\documentclass[aps,prd,preprint,showpacs,showkeys,nofootinbib,superscriptaddress,usenames,dvipsnames]{revtex4-1}

\usepackage{amsmath}
\usepackage{amssymb,slashed}
\usepackage{dcolumn}	
\usepackage{bm}			
\usepackage{bbm} 		
\usepackage{multirow}
\usepackage{blkarray}
\usepackage{tikz}
\usetikzlibrary{%
arrows,%
calc,%
fit,%
patterns,%
plotmarks,%
shapes.geometric,%
shapes.misc,%
shapes.symbols,%
shapes.arrows,%
shapes.callouts,%
backgrounds,%
chains,%
topaths,%
trees,%
petri,%
mindmap,%
matrix,%
folding,%
fadings,%
through,%
positioning,%
scopes,%
decorations.fractals,%
decorations.shapes,%
decorations.text,%
decorations.pathmorphing,%
decorations.pathreplacing,%
decorations.footprints,%
decorations.markings,%
shadows} 
\usepackage{ulem}
\usepackage{nicefrac}
\usepackage{feynmp}
\usepackage[]{xcolor}
\definecolor{hgreen}{rgb}{0,.3,0}
\definecolor{hred}{rgb}{.3,0,0}
\definecolor{hblue}{rgb}{0,0,.3}
\definecolor{LightGray}{gray}{0.95}
\usepackage[colorlinks=true,linkcolor=hblue,citecolor=hgreen,filecolor=hblue,
urlcolor=hred]{hyperref}
\makeatletter
\def\endfmffile{%
	\fmfcmd{\p@rcent\space the end.^^J%
		end.^^J%
		endinput;}%
	\if@fmfio
	\immediate\closeout\@outfmf
	\fi
	\ifnum\pdfshellescape>\z@
	\immediate\write18{mpost \thefmffile}%
	\fi}
\makeatother

\usepackage{hyperref,graphicx}           
\usepackage{pbox}
\usepackage{xcolor}
\usepackage[utf8]{inputenc}

\newcommand{\beq}{\begin{equation}}
\newcommand{\eeq}{\end{equation}}

\tikzset{ 
	scalar/.style={dashed},
	scalar-ch/.style={dashed,postaction={decorate},decoration={markings,mark=at
			position .55 with {\arrow{>}}}},
	fermion/.style={postaction={decorate}, 
	decoration={markings,mark=at
			position .55 with {\arrow{>}}}},
	plain/.style={},			
	gauge/.style={decorate, decoration={snake,segment 
	length=0.2cm}},
	gauge-na/.style={decorate, 
	decoration={coil,amplitude=4pt, segment
			length=5pt}}
}

\begin{document}
\preprint{\hbox{PITT-PACC-2502, UTWI-15-2025}}

\title{Cosmological Histories in Neutrino Portal Dark Matter}

\author{Amro E. B. Abdelrahim}
\affiliation{Pittsburgh Particle Physics, Astrophysics, and Cosmology Center, Department of Physics and Astronomy, University of Pittsburgh, Pittsburgh, USA}
\author{Brian Batell}
\affiliation{Pittsburgh Particle Physics, Astrophysics, and Cosmology Center, Department of Physics and Astronomy, University of Pittsburgh, Pittsburgh, USA}
\author{Joshua Berger}
\affiliation{Colorado State University, Fort Collins, CO 80523, USA}
\author{David McKeen}
\affiliation{TRIUMF, 4004 Wesbrook Mall, Vancouver, BC V6T 2A3, Canada}
\author{Barmak Shams Es Haghi}
\affiliation{Texas Center for Cosmology and Astroparticle Physics, Weinberg Institute for Theoretical Physics, Department of Physics, The University of Texas at Austin, Austin, TX 78712, USA}

\date{\today}
\begin{abstract}
We explore the diverse cosmological histories of a dark sector that is connected to the Standard Model (SM) via a Dirac sterile neutrino. 
The dark sector consists of a complex scalar and a Dirac fermion dark matter (DM) candidate protected by a global $U(1)$ stabilizing symmetry. 
Assuming the dark sector has negligible initial abundance and is populated from reactions in the SM thermal plasma during the radiation era, we show that the cosmological histories of the dark sector fall into four qualitatively distinct scenarios, each one characterized by the strengths of the portal couplings involving the sterile neutrino mediator. 
By solving Boltzmann equations, both semi-analytically and numerically, we explore these thermal histories and transitions between them in detail, including the time evolution of the temperature of the dark sector and the number densities of its ingredients.
We also discuss how these various histories may be probed by cosmology, direct detection, indirect detection, collider searches, and electroweak precision tests.
\end{abstract}



\maketitle

\section{Introduction}
\label{sec:intro}

The as-yet-unknown nature of dark matter (DM) remains one of the major unresolved problems in particle physics and cosmology. Although our current knowledge of DM is based on its gravitational interactions, the possibility of non-gravitational interactions of DM is strongly motivated in the light of the cosmological origin of DM and also with regard to prospects for DM searches. A broad range of phenomenological models utilize a thermal mechanism for establishing the DM abundance which, according to the strength of the non-gravitational interactions between DM and the Standard Model (SM), can be categorized into two main classes: freeze-out and freeze-in.  
These two categories indicate the thermal history of DM and the extent of its equilibration with the SM which varies from full chemical equilibrium (thermal freeze-out)~\cite{Lee:1977ua, Steigman:1984ac,Gondolo:1990dk}
to not reaching equilibrium at all (freeze-in)~\cite{Hall:2009bx}. An intriguing scenario, beyond simple models extending the SM by a single DM candidate,
is the possibility of DM residing within a 
dark sector that is connected to the visible sector 
through a portal interaction~\cite{Boehm:2003hm,Pospelov:2007mp,Feng:2008ya}.
Due to the complex internal dynamics of the dark sector, such scenarios can exhibit much richer thermal histories than the simple thermal freeze-out and freeze-in mechanisms.    

In this paper we present a detailed analysis of the cosmology and thermal history of a minimal dark sector interacting through the neutrino portal~\cite{Minkowski:1977sc, Yanagida:1979as, GellMann:1980vs, Glashow:1979nm, Mohapatra:1979ia, Schechter:1980gr}. The 
interaction Lagrangian for 
the dark sector is given by 
\begin{equation}
\mathcal{L} = - y\, \overline{L}_{L} \, H \, N_R - \lambda  \,
\phi^* \, \overline{N} \, \chi 
+ \textrm{h.c.}\,
\label{eq:modelLag}
\end{equation} 
The model contains a fermion mediator $N$ (a ``sterile neutrino''), a scalar field $\phi$, and a fermion $\chi$, with the latter serving as the DM candidate. Furthermore a portal coupling $y$ (dark coupling $\lambda$) couples the mediator to the $LH$ ($\phi \chi$) operators where $L$ and $H$ are the SM lepton and Higgs doublets. Our analysis is based on both semi-analytical and numerical solutions of the Boltzmann equations describing the time evolution of the dark sector number and energy densities. 
Focusing on dark sector masses near the TeV scale, the hierarchy $m_N < m_\chi < m_\phi$, and negligible initial dark sector abundance, we find four main phases regarding the thermal history of the dark sector and the mechanism that sets the final DM abundance: (i) {\it freeze-out}, which corresponds to large $y$ and large $\lambda$ such that the dark sector establishes chemical equilibrium with the SM bath and DM freezes out from it, (ii) {\it freeze-in}, that is characterized by large $y$ and small $\lambda$ so that $N$ stays in equilibrium with the SM while $\chi$ and $\phi$ do not, (iii) {\it double freeze-in}, which is marked by small $y$ and small $\lambda$ to such an extent that $N$ never reaches equilibrium with the SM, and finally (iv) {\it dark sector equilibration}, corresponding to small $y$ and large $\lambda$ so that the dark sector has a very weak coupling to the SM bath and a large dark interaction, leading to
formation of a distinct thermal bath in the dark sector with a temperature much smaller than the visible sector temperature. We analyse each region in detail, discuss their boundaries, and examine the evolution of the temperature of the dark sector and also the number densities of its constituents. Finally, we examine the existing constraints and future sensitivity projections imposed on the
neutrino portal dark sector by cosmology, indirect detection, colliders, direct detection, and electroweak precision data.  

Before proceeding to the main discussion, we briefly mention some of the relevant literature that complements our study. Various aspects of neutrino portal DM cosmology and phenomenology have been considered previously, see, e.g.,~Refs.~\cite{Pospelov:2007mp, Falkowski:2009yz, Kang:2010ha, Falkowski:2011xh, Cherry:2014xra, Bertoni:2014mva, Tang:2015coo, GonzalezMacias:2015rxl, Gonzalez-Macias:2016vxy,  Escudero:2016tzx, Escudero:2016ksa, Tang:2016sib, Campos:2017odj, Batell:2017rol, Batell:2017cmf, Schmaltz:2017oov, Folgado:2018qlv, Chianese:2018dsz,  Becker:2018rve, Berlin:2018ztp, Bandyopadhyay:2018qcv, Bertuzzo:2018ftf,  Kelly:2019wow, Blennow:2019fhy, Ballett:2019pyw, Chianese:2019epo, Patel:2019zky, Cosme:2020mck, Du:2020avz, Bandyopadhyay:2020qpn, Chianese:2020khl, Chianese:2021toe,  Biswas:2021kio, Kelly:2021mcd, Coito:2022kif, Liu:2022rst, Hepburn:2022pin, Barman:2022scg, Ahmed:2023vdb, Liu:2023kil, Liu:2023zah, Xu:2023xva, Das:2023yhv, Hong:2024zsn, Bell:2024uah} for some representative studies. Studies exploring diverse thermal histories within minimal dark sector portal models can be found in Refs.~\cite{Chu:2011be,Berlin:2016vnh,Heikinheimo:2016yds,Okawa:2016wrr,Berger:2018xyd,Hambye:2019dwd,Fitzpatrick:2020vba,Li:2022bpp}, while other novel thermal DM production mechanisms beyond freeze-out or freeze-in are discussed in Refs.~\cite{Griest:1990kh,Carlson:1992fn,Hochberg:2014dra,DAgnolo:2015ujb,Kuflik:2015isi,Pappadopulo:2016pkp,Dror:2016rxc,DAgnolo:2017dbv,Evans:2019vxr}.

The rest of our paper is organized as follows. In Section~\ref{sec:model}, we present the ingredients of our neutrino portal dark sector and provide an overview of its cosmology. Section~\ref{sec:cosmo}, which is the main part of this study, describes in detail the possible cosmologies for our model. The governing Boltzmann equations, with analytically-estimated solutions and numerical results, are used to map out 
regions of the parameter space corresponding to distinct dynamics responsible for the final abundance of the DM. 
In Section~\ref{sec:signals}, we examine the constraints and signatures of the model. Our conclusions are presented in Section~\ref{sec:discussion}. In Appendix~\ref{app:therm-avg}, we provide the relevant thermal averages of cross sections and decay rates. In particular, the results for thermally averaged cross section of colliding particles with different temperatures are presented for the first time.

\section{Neutrino Portal Model and Cosmology Overview}
\label{sec:model}
As a benchmark model for a dark sector connected to the SM 
via the neutrino 
portal, we supplement the SM matter content with three new 
fields: a Dirac 
sterile neutrino $N$, a Dirac fermion $\chi$, 
and a complex scalar 
$\phi$, with mass parameters $m_N$, $m_\chi$, and $m_\phi$, respectively. The $\chi$ and $\phi$ fields are assumed to be 
protected by a global 
$U(1)$ symmetry that stabilizes the DM\footnote{We do not address potential quantum gravity effects that may break this $U(1)$ symmetry, as such effects are generally model-dependent.}. The 
interaction Lagrangian for 
the dark sector is given by Eq.~(\ref{eq:modelLag}).
\begin{figure}[!tbh]
	\centering
	\begin{tikzpicture}
	\node at (0,0) {\begin{tikzpicture} 
		\draw[fermion] (0,0) node[left] {$L_L$} -- (1,-1);
		\draw[scalar-ch] (1,-1) -- (0,-2) node[left] {$H$};
		\draw[fermion] (1,-1) -- (2.5,-1) node[right] 
		{$N_R$};
	\end{tikzpicture}};
	\node at (6,0) {\begin{tikzpicture} 
	\draw[fermion] (0,0) node[left] {$t_R$} -- (1,-1);
	\draw[fermion] (1,-1) -- (0,-2) node[left] {$Q_L$};
	\draw[scalar-ch] (2.5,-1) -- node[above]{$H$} (1,-1);
	\draw[fermion] (2.5,-1) -- (3.5,0) node[right] {$N_R$};
	\draw[fermion] (3.5,-2) node[right] {$L_L$} -- (2.5,-1);
	\end{tikzpicture}};
	\node at (12,0) {\begin{tikzpicture} 
		\draw[fermion] (0,0) node[left] {$L_L$} -- (1,-1);
		\draw[scalar-ch] (1,-1) -- (0,-2) node[left] {$H$};
		\draw[fermion] (1,-1) -- node[above] {$N_R$} 
		(2.5,-1);
		\draw[scalar-ch] (2.5,-1) -- (3.5,-2) node[right] 
		{$\phi$};
		\draw[fermion] (2.5,-1) -- (3.5,0) node[right] 
		{$\chi$};
		\end{tikzpicture}};
	\end{tikzpicture}
	\caption{Processes contributing to the population of 
	the dark 
	sector. Two additional crossings contribute for the 
	second 
	diagram. }\label{fig:connect-diagrams}
\end{figure}
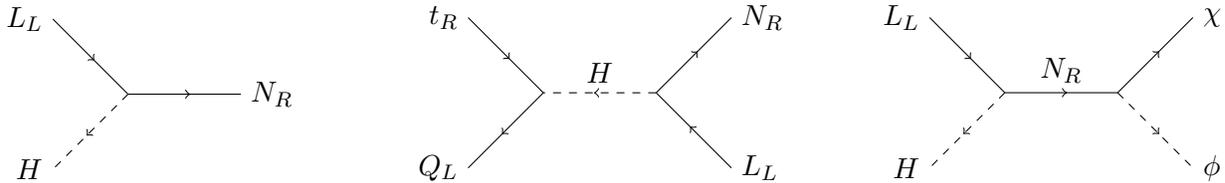
For concreteness, we focus on the 
interesting scenario in which the masses have the hierarchy
\begin{equation}
m_\phi > m_\chi > m_N > v\,, \label{eq:mass-hierarchy}
\end{equation}
where $v = 174~{\rm GeV}$ is the electroweak symmetry breaking scale.
With such a choice, $\chi$ serves as the DM candidate. A qualitatively similar cosmology and phenomenology would result if $\phi$ is lighter than $\chi$ (i.e., $m_\chi > m_\phi > m_N > v$) such that $\phi$ serves as the DM candidate. 
Furthermore, a qualitatively distinct scenario ensues if, instead of the hierarchy in Eq.~(\ref{eq:mass-hierarchy}),  $N$ is the heaviest state (e.g., $m_N > m_\phi > m_\chi$).
This is beyond our present scope, see, e.g., Refs.~\cite{Bandyopadhyay:2020qpn,Chianese:2020khl} for detailed studies. 

The cosmological setting for our analysis is the radiation era defined by temperature $T$ of the SM bath. For later use we define the Hubble parameter, energy density, and entropy density, 
\begin{equation}
\label{eq-defs-H-rho-s}
 H^2 = \frac{8 \pi \rho_{\rm tot} }{3 M_{\rm Pl}^2}, ~~~~~~~~  \rho_{\rm tot} = \frac{\pi^2}{30} g_{*}(T) T^4, ~~~~~~~~ s_{\rm tot} = \frac{2\pi^2}{45} g_{* S}(T) T^3,
\end{equation}
where $M_{\rm Pl} = 1.22 \times 10^{19}$ GeV.
For simplicity, we will fix the effective number of relativistic degrees of freedom to their SM values, $g_{*}  = g_{* S} = 106.75$ throughout our analysis. 
This is a good approximation since the relevant cosmological dynamics occurs at temperatures $T \gtrsim v$ and the number of degrees of freedom and fraction of energy density in the dark sector is small. 

In the limit 
of small $y$, only a small number of processes connect 
the SM sector to the 
dark sector. The (inverse) decay process, 
\begin{equation}
N \leftrightarrow \overline{H} + L\,,\label{eq:N decay reaction}
\end{equation}
is generally dominant at temperatures $T \sim m_N$.
At early times, the  $2 \leftrightarrow 2$ processes that produce one $N$ can dominate due to their different kinematics. Including only such processes that occur at 
tree-level and involve the 
relatively large top quark Yukawa coupling $y_t$,
we consider the 
process
\begin{equation}
t + \overline{Q} \leftrightarrow N + \overline{L}\,,\label{eq:N scattering}
\end{equation}
as well as its crossings. These processes populate $N$ and involve only the coupling $y$. The production of the other dark states has to involve the coupling $\lambda$, and is provided by the direct process
\begin{equation}
L + \overline{H} \leftrightarrow \chi + \overline{\phi}\,.\label{eq:LH chi phi}
\end{equation}
The diagrams for the relevant 
processes are shown in 
Fig.~\ref{fig:connect-diagrams}.  For simplicity, crossings of this process, which describe scattering of $\chi$ and $\phi$ off of bath particles are not included in our analysis. First of all, these processes taken together do not change the total number of $\chi$ and $\phi$ particles, which is the relevant quantity for estimating the DM abundance. Moreover, the processes taken individually have a similar effect to the decay process in~(\ref{eq: phi decay reaction}) below, and are subdominant in general. 

\begin{figure}[!tbh]
	\centering
	\begin{tikzpicture}
	\node at (9,-3.5) {\begin{tikzpicture} 
		\draw[fermion] (0,0) node[left] {$\chi$} -- (1,-1);
		\draw[fermion] (1,-1) -- (0,-2) node[left] {$N$};
		\draw[scalar-ch] (1,-1) -- (2.5,-1) node[right] 
		{$\phi$};
		\end{tikzpicture}};
	\node at (4.5,0) {\begin{tikzpicture} 
		\draw[fermion] (0,0) node[left] {$N$} -- (1.5,0);
		\draw[fermion] (1.5,0) -- (3,0) node[right] 
		{$\chi$};
		\draw[scalar-ch] (1.5,-2) -- node[right]{$\phi$} 
		(1.5,0);
		\draw[fermion] (3,-2)  node[right] {$\chi$} -- 
		(1.5,-2);
		\draw[fermion] (1.5,-2) -- (0,-2) node[left] {$N$} ;
		\end{tikzpicture}};
	\node at (9,0) {\begin{tikzpicture} 
	\draw[fermion] (0,0) node[left] {$N$} -- (1.5,0);
	\draw[scalar-ch] (3,0) node[right] {$\phi$}-- (1.5,0);
	\draw[fermion] (1.5,0) -- node[right]{$\chi$} (1.5,-2);
	\draw[scalar-ch] (1.5,-2) -- (3,-2)   node[right] 
	{$\phi$};
	\draw[fermion] (1.5,-2) -- (0,-2) node[left] {$N$} ;
	\end{tikzpicture}};
	\node at (13.5,0) {\begin{tikzpicture} 
		\draw[fermion] (0,0) node[left] {$\chi$} -- (1.5,0);
	\draw[scalar-ch] (3,0) node[right] {$\phi$}-- (1.5,0);
	\draw[fermion] (1.5,0) -- node[right]{$N$} (1.5,-2);
	\draw[scalar-ch] (1.5,-2) -- (3,-2)   node[right] 
	{$\phi$};
	\draw[fermion] (1.5,-2) -- (0,-2) node[left] {$\chi$} ;
	\end{tikzpicture}};
\end{tikzpicture}
\caption{Internal dark sector interactions including annihilation, and decay - inverse decay processes. 
}\label{fig:dynamics-diagrams}
\end{figure}
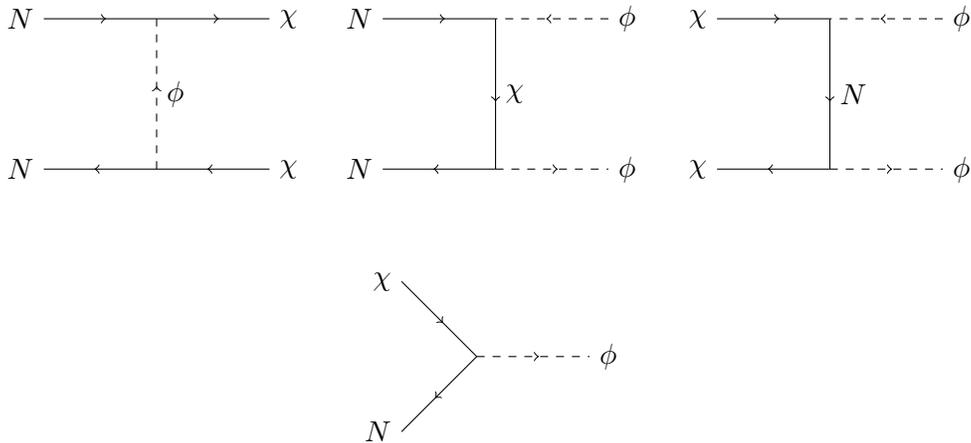
Within the dark sector, we make the simplifying assumption that the phase space distributions follow the equilibrium form with 
common temperature $T^\prime$. This should provide a good approximation to the true distributions for the cases studied here. 
If the couplings are large enough, kinetic equilibrium is reached in the dark sector. Even when this is not the case, the typical momentum of the dark sector particles produced from the SM bath is of order the SM temperature $T' \sim T$. The  cosmological dynamics of 
the dark sector are then described by four quantities which 
we can take to be 
the dark sector temperature $T^\prime$ along with the dark 
sector co-moving number densities (or ``yields'') $Y_N$, 
$Y_\chi$, and $Y_\phi$, with the usual definition
\begin{equation}
\label{eq-yield-def}
Y_i = \frac{n_i}{s_{\rm tot}}\,,
\end{equation} 
where $n_i$ is the number density of species $i$ and $s_{\rm tot}$ is 
the entropy density (\ref{eq-defs-H-rho-s}). Since the energy in the dark sector is 
conserved up to its 
interactions with the SM, the temperature is determined by 
the connecting 
interactions in Fig.~\ref{fig:connect-diagrams}, along with the 
standard statistical 
determination of the energy density in terms of 
temperature. The dark yields are determined by their number-changing annihilation processes
\begin{equation}
    N+\overline{N}\leftrightarrow \chi+\overline{\chi}\,,\label{eq:NN chi chi}
\end{equation}
\begin{equation}
    N+\overline{N}\leftrightarrow \phi+\overline{\phi}\,,\label{eq:NN phi phi}
\end{equation}
\begin{equation}
    \chi+\overline{\chi}\leftrightarrow \phi+\overline{\phi}\,,\label{eq: chi chi phi phi}
\end{equation}
together with the $\phi$ (inverse) decay process
\begin{equation}
    \phi\leftrightarrow\chi +\overline{N}\,.\label{eq: phi decay reaction}
\end{equation}
The diagrams for these internal dark processes are shown in Fig.~\ref{fig:dynamics-diagrams}.

We also assume that the initial density of dark sector particles is vanishing or negligible, so that they are created from the SM bath. However, not all dark particles are created equal: $N$ occupies the distinct position 
of interacting
directly with the SM particles via the coupling $y$, whereas $\chi$ and $\phi$ interactions with the SM are mediated by $N$ and thus require 
both couplings $y$ and $\lambda$. This situation leads to a division between the dynamics of population of the dark sector and the dynamics within the dark sector. 

 First, there is a division that depends on the value of $y$. If this coupling is large enough, $N$ would reach equilibrium with the SM. In this case, $N$ is in effect part of the SM bath, and thus DM couples directly to the bath, which is the familiar situation in DM models employing, e.g., freeze-out or freeze-in production. 
 If instead $y$ is small enough, $N$ is effectively part of the dark sector.
 
 The strength of the coupling $\lambda$ determines the dynamics within the dark sector and hence the creation mechanism of DM. For small values of $\lambda$, $N$ has no effect, other than as a mediator to the SM, on $\chi$ and $\phi$, and their evolution is decoupled. The important processes are $L+\overline{H}\rightarrow \chi + \overline{\phi}$ and $\phi\rightarrow \chi+\overline{N}$. As $\lambda$ is dialed up, the various dark processes begin to become relevant starting with scatterings $N+\overline{N}\rightarrow \chi+\overline{\chi}, \phi+\overline{\phi}$ as $N$ is most abundant, then the inverse decay $\overline{N}+\chi \rightarrow \phi$, and ultimately all of the processes in Eqs.~(\ref{eq:NN chi chi},\ref{eq:NN phi phi},\ref{eq: chi chi phi phi},\ref{eq: phi decay reaction}) can play a role with increasing $\lambda$. 

Perhaps the most novel case is that of small $y$, large $\lambda$, which we now discuss. As we are assuming a common dark sector temperature $T'$, 
the remaining 
question is then whether (i) the dark sector reaches chemical 
equilibrium with 
$\mu_N = \mu_\chi = \mu_\phi = 0$, (ii) reaches partial chemical 
equilibrium with 
$\mu_N = \mu_\chi \equiv \mu \neq 0$, or (iii) remains in 
chemical disequilibrium. 
In the first scenario, the dark sector population drastically cools with respect to the SM as the energy density of the dark sector is divided 
among a large number of particles. With our mass hierarchy (\ref{eq:mass-hierarchy}), we find that this first scenario invariably transitions into the second scenario as the dark temperature falls below the mass of the heaviest dark particle $\phi$. This second scenario of partial $\chi-N$ equilibrium unfolds as a more 
standard freeze out, with the dark sector temperature $T^\prime \lesssim T$. The final scenario is the freeze-in limit. 

The applicable scenario depends on the 
strength of 
number-changing processes in the dark sector. Full 
chemical equilibrium 
is reached if the $\phi$ (inverse) decay process (\ref{eq: phi decay reaction}) reaches equilibrium, as it is the only process within the dark sector that changes total dark particle number. Note that a $2 \to 2$ process 
such as (\ref{eq:NN chi chi}) or (\ref{eq:NN phi phi}) must also have a 
sufficiently large rate to 
populate the species of the dark sector other than $N$ and 
``turn on'' the 
inverse decay process. The two processes initiate the approach to equilibrium, which is fully reached only when a third process, typically (\ref{eq:NN phi phi}), reaches equilibrium. Although the dark sector has reached equilibrium, the production from the SM is still active, producing mainly $N$s (which have a temperature of the order of the visible sector temperature) that subsequently equilibrate. The dark sector is thus heated relative to the visible sector with the dark temperature still decreasing, albeit at a slower rate because of the heating. With the hierarchy 
that we consider, $T^\prime$ crosses below $m_\phi$ first and the $\phi$ annihilation
processes go out of equilibrium, while the $2 \to 2$ 
process (\ref{eq:NN chi chi}) remains effective. 
Although general wisdom suggests that the reaction $N + \overline{N} 
\rightarrow \chi + \overline{\chi}$ becomes Boltzmann-suppressed after DM becomes non-relativistic, the fact that there is still active production of energetic $N$s from the SM 
counteracts the Boltzmann suppression, keeping the dark temperature and DM density constant. As the universe expands, the produced $N$s do not have enough energy to keep the dark temperature constant, hence the Boltzmann suppression of DM density resumes, leading eventually to $\chi$ annihilation, freeze-out, and the creation of a relic abundance 
of $\chi$. 
If $\lambda$ is very small, then none of these 
processes ever enter 
equilibrium.

The division between the dynamics of the population of the 
dark sector and the 
dynamics within the dark sector naturally leads 
to four qualitatively 
different potential cosmological histories, which we 
describe in detail below. 

\section{Cosmological histories}
\label{sec:cosmo}
 In this section, we delve into the possible cosmologies for 
the model we consider.  
We begin by setting up the Boltzmann equations 
that we will use to 
study the evolution of the energy and number densities. We then develop approximate solutions in several limiting regimes to 
determine the rough boundaries of the qualitatively 
different behaviors that 
are possible. Finally, we present our numerical results for 
the full solution 
to these Boltzmann equations for several benchmark cases in 
which the dark 
matter relic abundance matches observation.

To simplify our analysis, we work in the good approximation that the dark 
sector follows a thermal distribution with a single temperature $T^\prime$ over 
all the parameter space that we consider. This is certainly justified in the 
regime where $\lambda$ is sufficiently large to thermalize the dark sector. In the small $\lambda$ regime, the dark sector has a non-thermal distribution.  If dark number changing processes that go like $\lambda$ within the dark sector are sufficiently fast, then the kinetic energy that the dark sector particles acquire of $\mathcal{O}(T)$ can be divided among many new particles. The chemical potential goes to zero and the temperature of the dark sector drops to $T^\prime \ll T$.  On the other hand, when $\lambda$ is small, the dark sector has a significant chemical potential and remains with kinetic energy of $\mathcal{O}(T)$ in a non-thermal distribution.  In this small $\lambda$ regime, dark number changing processes are not in 
equilibrium by dividing its energy among a large number of 
particles. The average kinetic energy of dark sector particles therefore 
remains of order $T$. While the distribution is not exactly Maxwell-Boltzmann,  
we expect that the distortion of the distribution has only a small effect on the bulk quantities of 
interest. 

\subsection{Boltzmann Equations}
Energy is conserved within the dark sector and we assume 
dark sector 
kinetic equilibrium, so there is only one relevant equation 
aside from those for the 
abundances of the various species. We can take that to be 
the total dark sector energy density equation,
\begin{eqnarray}\label{eq:energy-boltz}
\frac{d\rho}{dt} + 3 H (\rho + p) & = & 
\langle \Gamma_{N,{\rm tot}} E_N  \rangle_T \, n^{\rm eq}_{N}(T) 
-  \langle \Gamma_{N,{\rm tot}} E_N  \rangle_{T^\prime}\, n_N   
\nonumber\\& + &
\langle (\sigma v)_{\chi\phi LH} (E_\chi + E_\phi) \rangle_T \, n_{\chi}^{\rm eq}(T) \, n_{\phi}^{\rm eq}(T) 
- \langle (\sigma v)_{\chi\phi LH}  (E_\chi + E_\phi) \rangle_{T^\prime} \, n_\chi \, n_\phi, ~~~~~~
\end{eqnarray}  
where $\rho = \rho_N + \rho_\chi + \rho_\phi$ is the total dark 
sector energy 
density, $p = p_N + p_\chi + p_\phi$ is the total dark 
sector pressure,  $H$ in the second term on the l.h.s is the Hubble parameter defined in Eq.~(\ref{eq-defs-H-rho-s}),
$\langle \Gamma_{N,{\rm tot}}E_N\rangle$ is the thermally-averaged total rate of 
energy exchange between the two sectors 
for the processes involving one $N$, 
i.e. the (inverse) decay and scatterings (\ref{eq:N decay reaction},\ref{eq:N scattering}), $\langle  (\sigma 
v)_{\chi\phi LH} (E_\chi 
+ E_\phi) \rangle$ is the thermally-averaged cross section times energy 
for $\chi$ and $\phi$ to 
annihilate into SM particles through the process $\chi + \overline \phi \rightarrow L + \overline H$, 
and $n_{i}^{\rm eq}(T)$ 
denotes the number 
density for species $i$ when it is in chemical equilibrium 
at a temperature 
$T$. Note that $\langle\Gamma_{N,{\rm tot}}E_N\rangle_{T^\prime}$ contains scattering
of $N$, which we assume has temperature $T^\prime$, off of a bath particle,
whose temperature is $T$. The thermally-averaged cross section times
velocity for the scattering of two particles at different temperatures is reduced to
a single integral of the Gondolo-Gelmini type~\cite{Gondolo:1990dk} in Appendix \ref{app:therm-avg}. 

While the total dark sector energy density is affected only by the connecting
interactions shown in Fig.~\ref{fig:connect-diagrams}, the number density of $N$ is determined by both
its interactions with the SM bath (\ref{eq:N decay reaction},\ref{eq:N scattering}), its annihilation to dark sector
particles (\ref{eq:NN chi chi},\ref{eq:NN phi phi}), and $\phi$ (inverse) decays (\ref{eq: phi decay reaction}),
shown in Fig.~\ref{fig:dynamics-diagrams}, and its evolution is given by: 
\begin{align} 
\label{eq:N yield-boltz}
\frac{dn_{N}}{dt}+3 H n_{N} & = 
\langle\Gamma_{N,{\rm tot}}\rangle_{T} \, n_{N}^{\rm eq}(T)
-\langle\Gamma_{N,{\rm tot}}\rangle_{T^\prime} \, n_{N}
+\langle\Gamma_{\phi}\rangle_{T^\prime}\left[n_{\phi}
-n_{\phi}^{\rm eq}(T^\prime)\frac{n_{\chi}}{n_{\chi}^{\rm eq}(T^\prime)}\frac{n_{N}}{n_{N}^{\rm eq}(T^\prime)}\right] \nonumber \\
& +  \langle(\sigma v)_{\chi\chi NN}\rangle_{T^\prime}\left[n_{\chi}^{2}
-\frac{n_{\chi}^{\rm eq}(T^\prime)^2}{n_{N}^{\rm eq}(T^\prime)^2} \, n_{N}^{2}\right]
+\langle(\sigma v)_{\phi\phi NN}\rangle_{T^\prime}\left[n_{\phi}^{2}
-\frac{n_{\phi}^{\rm eq}(T^\prime)^2}{n_{N}^{\rm eq}(T^\prime)^2} \, n_{N}^{2}\right].
\end{align}
Similar equations hold for $\chi$ and $\phi$ number densities: 
\begin{align}
\label{eq:chi yield-boltz}
\frac{dn_{\chi}}{dt}+3 H n_{\chi} & = 
\langle(\sigma v)_{\chi\phi LH}\rangle_{T} \, n_{\chi}^{\rm eq}(T) \, n_{\phi}^{\rm eq}(T)
-\langle(\sigma v)_{\chi\phi LH}\rangle_{T^\prime} \, n_{\chi} \, n_{\phi}
\nonumber \\ & +  
\langle\Gamma_{\phi} \rangle_{T^\prime}\left[n_{\phi}
-n_{\phi}^{\rm eq}(T^\prime)\frac{n_{\chi}}{n_{\chi}^{\rm eq}(T^\prime)}\frac{n_{N}}{n_{N}^{\rm eq}(T^\prime)}\right]
-\langle(\sigma v)_{\chi\chi NN}\rangle_{T^\prime}\left[n_{\chi}^{2}
-\frac{n_{\chi}^{\rm eq}(T^\prime)^2}{n_{N}^{\rm eq}(T^\prime)^{2}}\, n_{N}^{2}\right]
\nonumber \\ & + 
\langle(\sigma v)_{\phi\phi\chi\chi}\rangle_{T^\prime}\left[n_{\phi}^{2}
-\frac{n_{\phi}^{\rm eq}(T^\prime)^2}{n_{\chi}^{\rm eq}(T^\prime)^2} \, n_{\chi}^{2}\right],
\end{align}
and
\begin{eqnarray}\label{eq:phi yield-boltz}
\frac{dn_{\phi}}{dt}+3 H n_{\phi} & = & 
\langle(\sigma v)_{\chi\phi LH}\rangle_{T} \, n_{\chi}^{{\rm eq}}(T) \, n_{\phi}^{{\rm eq}}(T)
-\langle(\sigma v)_{\chi\phi LH}\rangle_{T^\prime} \, n_{\chi} \, n_{\phi}\nonumber 
\\ & &
-\langle\Gamma_{\phi}\rangle_{T^\prime}\left[n_{\phi}
-n_{\phi}^{{\rm eq}}(T^\prime)\frac{n_{\chi}}{n_{\chi}^{{\rm eq}}(T^\prime)}\frac{n_{N}}{n_{N}^{{\rm eq}}(T^\prime)}\right]
-\langle(\sigma v)_{\phi\phi NN}\rangle_{T^\prime}\left[n_{\phi}^{2}
-\frac{n_{\phi}^{{\rm eq}}(T^\prime)^2}{n_{N}^{{\rm eq}}(T^\prime)^2} \, n_{N}^{2}\right]
\nonumber \\ & &
-\langle(\sigma v)_{\phi\phi\chi\chi}\rangle_{T^\prime}\left[n_{\phi}^{2}
-\frac{n_{\phi}^{{\rm eq}}(T^\prime)^2}{n_{\chi}^{{\rm eq}}(T^\prime)^2} \, n_{\chi}^{2}\right],
\end{eqnarray}
where $\langle\Gamma_{N,{\rm tot}}\rangle_{T}$ is the thermally-averaged total rate of $N$ production by inverse decays and scatterings of SM particles, $\langle\Gamma_{\phi}\rangle_{T^\prime}$ is the thermally-averaged decay rate for the process $\phi \rightarrow \chi + \overline N$,
while $\langle(\sigma v)_{\chi\chi NN}\rangle_{T^\prime}$, $\langle(\sigma v)_{\phi\phi NN}\rangle_{T^\prime}$, $\langle(\sigma v)_{\phi\phi\chi\chi}\rangle_{T^\prime}$
represent the thermally-averaged cross section times velocity for
the annihilation processes $\chi + \overline{\chi}\rightarrow N + \overline{N}$,
$\phi + \overline{\phi}\rightarrow N + \overline{N}$ and $\phi + \overline{\phi}\rightarrow\chi + \overline{\chi}$
respectively. Our notational conventions along with detailed expressions for the various cross sections,
decay rates, and their thermal averages are given in Appendix~\ref{app:therm-avg}. 

Assuming a negligible or vanishing initial density of dark sector
particles, the connecting processes shown in Fig.~\ref{fig:connect-diagrams} serve to produce
them. The evolution divides into an early-time production phase and a late-time behavior governed by how large the dark coupling $\lambda$ is. We will investigate the early-time behavior first.

\subsection{Evolution at Early times}
At early times, before a large abundance of dark
particles had time to build up and when all particles are relativistic, the behavior is independent of the dark sector density and temperature and depends only on the SM injection. One can, therefore, neglect all terms in the above
Boltzmann equations except those producing dark particles from the
SM bath. These terms are given functions and can be integrated to get early-time solutions for all unknowns. For $N$, 
Eq.~(\ref{eq:N yield-boltz}) gives: 
\begin{equation}
\label{eq-N-prod}
Y_{N}=\frac{g_{N}\Gamma_{N}m_{N}^{3}}{6\pi^{2}H(m_{N})s(m_{N})}x^{3}+\frac{g_{\rm SM}g_{N}\langle(\sigma v)_{N,{\rm SM}}\rangle_{T = m_N}^{\rm rel}m_{N}^{6}}{\pi^{4}H(m_{N})s(m_{N})}x,
\end{equation}
where $\Gamma_{N}\simeq\frac{y^{2}m_{N}}{8\pi g_{N}}$ 
is the
decay rate of $N$, $\langle(\sigma v)_{N,{\rm SM}}\rangle_{T}^{\rm rel}\approx\frac{9y_{t}^{2}y^{2}}{64\pi g_{\rm SM}g_{N}T^{2}}$
is the relativistic approximation of the thermally-averaged effective cross section of processes (4) describing $N$ production by scattering
of SM bath particles, and $g_N =2$ ($g_{\rm SM} = 1$) represents the number of internal degrees of freedom of $N$ (single massless SM degree of freedom). Note that we have changed variables from number
density to the conventional yield given by Eq.~(\ref{eq-yield-def}), and from time $t$ to
$x=m_{N}/T$
using Eq.~(\ref{eq-defs-H-rho-s}).
$N$ production by scattering of bath particles 
-- the second term in Eq.~(\ref{eq-N-prod}) -- dominates over the inverse decay early
on (until $x \sim 1$) as indicated by the linear vs cubic dependence on $x$. For convenience,
we recast this equation as 
\begin{equation}
\label{eq: early N yield}
Y_{N}=Y_{N}^{\rm eq, rel}\left(\frac{x^{3}}{3x_{{\rm id},N}^{3}}+\frac{x}{x_{{\rm scat},N}}\right)\approx Y_{N}^{\rm eq, rel} \frac{x}{x_{{\rm scat},N}},
\end{equation}
where $Y_{N}^{\rm eq, rel}=\frac{g_{N}}{\pi^{2}}\frac{m_{N}^{3}}{s(m_{N})}$ is
the equilibrium yield when the particle is relativistic. The inverse temperatures
$x_{{\rm id},N}=\left(2H(m_{N})/\Gamma_{N}\right)^{1/3}$, 
$x_{{\rm scat},N}=H(m_{N})/(\frac{g_{\rm SM}}{\pi^{2}}m_{N}^{3}\langle(\sigma v)_{N,SM}\rangle_{T=m_N}^{\rm rel})$
are the respective times at which the inverse decay rate and scattering rate
of $N$ become equal to the Hubble rate. This expression allows us
to determine values of $y$ for which $N$ equilibrates with the
SM before a particular time $x_{\rm DM}$ related to DM dynamics (e.g., for direct freeze-in of DM via the process $L+\overline{H}\rightarrow \chi+\overline{\phi}$, 
we have $x_{\rm DM} = x_{\rm fin} \sim m_N/m_{\phi}$):
\begin{equation}\label{eq:N equili condition}
y \gtrsim \left(\frac{64\pi^{3}g_{N}}{9y_{t}^{2}}\frac{H(m_{N})}{m_{N}}\frac{1}{x_{\rm DM}}\right)^{1/2} \simeq 1.3\times10^{-6}\left(\frac{{m_{N}}/{m_{\phi}}}{x_{\rm DM}}\right)^{1/2}\left(\frac{m_{\phi}}{3 ~ {\rm TeV}}\right)^{1/2},
\end{equation}
and this separates the $y-\lambda$ parameter space into a  
large $y$ region where $N$ thermalizes with the SM bath at early times and a small $y$ region where it does not.  

As for the early-time yields of $\chi$ and $\phi$, Eqs.~(\ref{eq:chi yield-boltz},\ref{eq:phi yield-boltz}) imply
\begin{equation}\label{eq:early chi yield}
Y_{\chi}=Y_{\phi}=\frac{g_{\chi}g_{\phi}\langle(\sigma v)_{\chi\phi LH}\rangle_{T=m_N}^{\rm rel}m_{N}^{6}}{\pi^{4}H(m_{N})s(m_{N})}x=Y_{\chi}^{\rm eq, rel}\frac{x}{x_{LH}},
\end{equation}
where $\langle(\sigma v)_{\chi\phi LH}\rangle_{T}^{\rm rel}\approx\frac{\lambda^{2}y^{2}}{128\pi g_{\chi}g_{\phi}T^{2}}$, and 
 $x_{LH}=H(m_{N})/(\frac{g_{\phi}}{\pi^{2}}m_{N}^{3}\langle(\sigma v)_{\chi\phi LH}\rangle_{T=m_N}^{\rm rel})$ 
is the time at which the rate of the process $L + \overline{H}\rightarrow\chi+ \overline{\phi}$
becomes equal to the Hubble rate. If $x_{LH} > x_{\rm fin}\sim m_{N}/m_{\phi}$, 
then the process $L + \overline{H}\rightarrow\chi+ \overline \phi$ becomes inactive before producing an equilibrium density of $\chi$ and one gets a freeze-in production of DM. 
This occurs when the product of couplings satisfies the condition  
\begin{equation}\label{eq: chi fin cond}
\lambda \, y \lesssim \left(128\pi^{3}g_{\chi}\frac{H(m_{N})}{m_{N}}\frac{1}{{m_{N}}/{m_{\phi}}}\right)^{1/2} \simeq 5.8\times10^{-6}\left(\frac{m_{\phi}}{3~{\rm TeV}}\right)^{1/2}.
\end{equation}

Finally, an integration of Eq.~(\ref{eq:energy-boltz}) yields
\begin{equation}
x^{4}\rho=2m_{N}s(m_{N})\left(Y_{N}^{\rm eq, rel}\left(\frac{x^{3}}{3x_{{\rm id},N}^{3}}+\frac{x}{x_{{\rm scat},N}}\right)+2Y_{\chi}^{\rm eq, rel}\frac{x}{x_{LH}}\right).
\end{equation}
The dark sector energy density at early times is given by: $\rho=3P=3(n_{N}+n_{\chi}+n_{\phi})T'=3sY_{\rm{tot}}T'$,
which leads to
\begin{equation}
\label{eq:TPdivT-1}
\frac{T'}{T}=\frac{2}{3}\frac{1}{Y_{\rm{tot}}}\left(Y_{N}^{\rm eq, rel}\left(\frac{x^{3}}{3x_{{\rm id},N}^{3}}+\frac{x}{x_{{\rm scat},N}}\right)+2Y_{\chi}^{\rm eq,rel}\frac{x}{x_{LH}}\right).
\end{equation}
This equation holds for most of the parameter space, except when the
back-reaction terms are large and drive the dark temperature to equal
the visible temperature. In the freeze-in limit, the total yield is
dominated by $N$ ($\lambda$ is not large enough to substantially
convert the $N$ density into $\chi$ and $\phi$) and the term in
brackets is the total yield, and thus one gets:
\begin{equation}
\label{eq:early temp}
T'=\frac{2}{3}T,
\end{equation}
which, as mentioned earlier, is the freeze-in result for the temperature;
the dark particles cannot divide the energy amongst themselves (feeble
interaction) and therefore their temperature remains of the order
of the SM temperature. 

\begin{figure}[ht]
\includegraphics[width=0.65\textwidth]{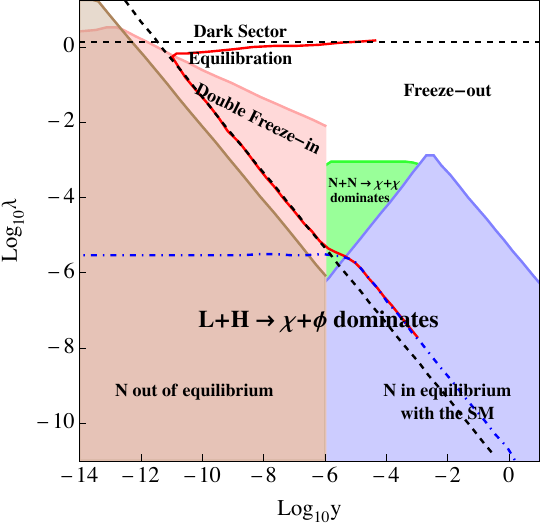} 
\caption{{\it Qualitative regimes of the parameter space:} the possible creation mechanisms of DM in our model as regions in the ($ y-\lambda$) parameter space with $m_N=200~{\rm GeV}$, $m_{\chi}=2~{\rm TeV}$, $m_{\phi}=3~{\rm TeV}$. The vertical line is given by Eq.~(\ref{eq:N equili condition}), and it determines whether $N$ is in equilibrium with the SM bath. The other boundaries between the different regions are described in the text. In the colored regions, DM is produced by freeze-in via either of the two competing processes $L+\overline{H}\rightarrow \chi+\overline{\phi}, N+\overline{N}\rightarrow \chi+\overline{\chi}$, with the first one dominating in the small $\lambda$ regime. For very large values of $\lambda$, DM is produced by freeze-out of $\chi+\overline{\chi}\rightarrow N+\overline{N}$, with $N$ being in or out of equilibrium (freeze-out and dark sector equilibration regions respectively). Along the solid red line, the DM abundance matches the observed value today. While this red contour is obtained numerically, the various dashed lines represent approximate analytical solutions to the Boltzmann equations, derived in the text.}
\label{fig:paramsummary}
\end{figure}

The subsequent behavior of the system depends on the strength of the
coupling to the visible sector $y$ and the dark coupling $\lambda$, 
which determine how fast the dark sector density builds up. When $\lambda$
is very small, dark processes are not fast and the DM abundance is determined
by the freeze-in of the SM production process $L+\overline{H}\rightarrow\chi+ \overline{\phi}$
(together with the decay $\phi\rightarrow\overline{N}+ \chi $). As one increases
$\lambda$, conversion of $N$ (the most abundant dark particle since
production proceeds via the large top's Yukawa coupling) into $\chi$
and $\phi$ starts to dominate DM production. According to whether
$N$ is in equilibrium with the SM or not, as given by Eq.~(\ref{eq:N equili condition}), we get
a freeze-in or a double freeze-in scenario for the process $N+\overline{N}\rightarrow\chi+\overline{\chi},\phi+\overline{\phi}$.
Increasing $\lambda$ further results in the dark sector establishing thermal equilibrium,
either with itself (for small $y$) leading to the dark thermalization
scenario described earlier, or with the SM giving the usual WIMP freeze-out. In the following subsections, we investigate each regime in detail; Fig.~\ref{fig:paramsummary} labels these regimes in the $y$-$\lambda$ parameter space, and compares the numerical solution of our Boltzmann system of equations with the semi-analytical results to be discussed next.

\subsection{Freeze-in}
This regime of standard freeze-in is characterized  by $N$ being in equilibrium with the SM (large $y$), while $\chi$ and $\phi$ are not (small $\lambda$). Thus DM particles are frozen in from the SM bath (blue and green regions in Fig.~\ref{fig:paramsummary}). There are two competing processes: $L+\overline {H}\rightarrow\chi+ \overline{\phi}$ and $N+\overline{N}\rightarrow\chi+\overline{\chi}$. Since $\phi$ would eventually
dump its density into $\chi$ via the decay process $\phi\rightarrow\overline{N}+\chi$, one also needs to include the process $N+\overline{N}\rightarrow\phi+\overline{\phi}$, and hence
the correct variable to follow is the total $\phi+\chi$ yield $Y_{\phi+\chi}$. The back scattering terms are negligible,
and our system of equations simplifies to
\begin{align}
\label{eq:phi+chi yield}
Hsx \frac{dY_{\phi+\chi}}{dx}  = 2\langle(\sigma v)_{\chi\phi LH}\rangle_{T} \, n_{\chi}^{\rm eq}(T) n_{\phi}^{\rm eq}(T)
&+\langle(\sigma v)_{\chi\chi NN}\rangle_{T^\prime}\frac{n_{\chi}^{\rm eq}(T^\prime)^2}{n_{N}^{\rm eq}(T^\prime)^2}n_{N}^{2}\nonumber \\ &+\langle(\sigma v)_{\phi\phi NN}\rangle_{T^\prime}\frac{n_{\phi}^{\rm eq}(T^\prime)^2}{n_{N}^{\rm eq}(T^\prime)^2}n_{N}^{2}.
\end{align}
When the process $L+ \overline{H}\rightarrow\chi+ \overline{\phi}$ dominates (first term on RHS of Eq.~(\ref{eq:phi+chi yield})), we are back to the early-time behavior described in Eq.~(\ref{eq:early chi yield}). As the temperature falls below the mass of $\phi$, this process shuts off producing the following DM relic abundance: 
\begin{equation}
\label{eq:freeze-in-LH-xfin}
Y_{\chi}^{\rm fin}=Y_{\phi+\chi,\infty}\simeq 2 \frac{Y_{\chi}^{\rm eq,rel}}{x_{LH}}x_{\rm fin}=c_{1}\frac{\lambda ^{2} y^{2}m_{N}^{5}}{64\pi^{5}H(m_{N})s(m_{N})m_{\phi}},
\end{equation}
where $c_1$ is an order one numerical factor that depends on $m_{\phi}$ ($c_1 \approx 1.06$ for $m_{\phi}=3~\rm{{TeV}}$). This expression for the yield holds in both the blue and brown regions in Fig.~\ref{fig:paramsummary}, however DM is under-produced in the brown region, and the correct relic abundance can only be obtained in the blue region. Since $Y_{\chi}^{\rm fin}\propto y^2\lambda^{2}$, the contour of correct DM relic abundance in the blue region is a straight line of negative slope.  The boundary between this region and the freeze-out region (rightmost line separating blue and white parts in Fig. \ref{fig:paramsummary} is given in Eq. \ref{eq: chi fin cond}.

Note that the freeze-in time $x_{\rm fin}= c_{1} m_N/m_{\phi}$ is not the only important time scale here. There is also the decay time $x_{\rm {d},\phi}\sim \sqrt{H(m_{N})/\Gamma_{\phi}}$ associated with the $\phi$ decay process, so that the final relic abundance is produced at time $x= {\rm max}[x_{\rm fin},x_{{\rm d},\phi}]$. 
Moreover, the inverse decay process becomes important as $\lambda$ is dialed up, and so the decay could also be an in-equilibrium decay. While such internal dark sector dynamics does not affect our expression for the relic abundance, it is interesting in its own right. A more thorough treatment can be found in the thesis~\cite{ElDawBaboAbdelrahim:2023rll}.

\begin{figure}[ht]
\centering
\includegraphics[width=0.49\textwidth]{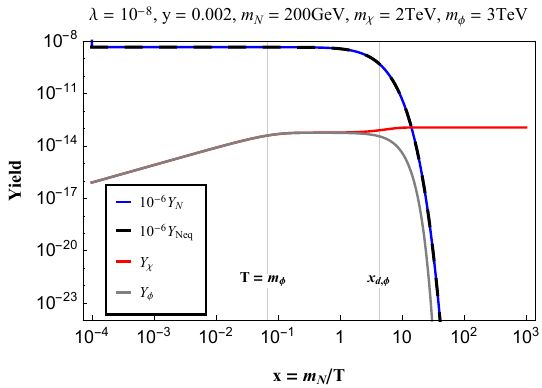}~~
\includegraphics[width=0.49\textwidth]{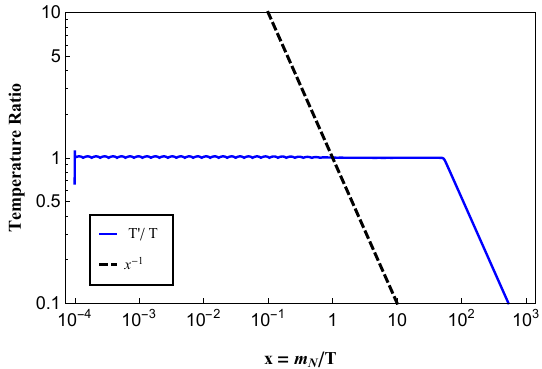}
\caption{The dark sector yields (left panel) and their temperature relative
to the visible temperature (right panel) in the standard freeze-in
regime. The coupling $y$, of $N$ to the SM particles, is large and
leads $N$ to equilibrate with the SM as the blue line and dashed
black one indicate. The coupling $\lambda$ is so small that internal
dark sector processes are not important and the process $L+\overline{H}\rightarrow\chi+\overline{\phi}$ dominates
DM production. This manifests as equal yields for $\chi$ and $\phi$
and the processes freeze in when the temperature falls below the $\phi$
mass (horizontal grid line on the left). Around the time $x_{\rm {d},\phi}$, $\phi$
starts to decay, increasing the DM density by a factor of 2 and producing
its final relic abundance. Note that $N$ is the most abundant dark
particles (its yield is scaled down by a factor of $10^{6}$ to appear
on the plot) and hence controls the behavior of the dark temperature,
leading it to equal the visible temperature. At late times, $N$ becomes
non-relativistic and the dark temperature decreases as $x^{-2}$ as
can be inferred from the black dashed line. Both $N$ and the dark temperature have no effect on DM due to $\lambda$ being small.}
\label{fig: standard freeze-in}
\end{figure}

When the other processes ($N+\overline{N}\rightarrow\chi+\overline{\chi}/\phi+\overline{\phi}$)
dominate, one can rewrite the last two terms in Eq.~(\ref{eq:phi+chi yield}) and integrate (using the fact that $T'=T$ in this regime) to obtain 
\begin{equation}
\label{eq:freeze-in-NN-x}
Y_{\phi+\chi}=Y_{\chi}^ {\rm eq, rel}\frac{x}{x_{N\chi}}+Y_{\chi}^{\rm eq, rel}\frac{x}{x_{N\chi}}\left[1-\gamma+\ln\left(\frac{2 \, m_{N}}{x\, m_{\chi}}\right)\right],
\end{equation}
where $x_{N\chi}=H(m_{N})/(\frac{g_{\chi}}{\pi^{2}}m_{N}^{3}\langle(\sigma v)_{\chi\chi NN}\rangle_{T=m_N}^{\rm rel})$ is the time at which the Hubble rate equals the rate of the process
$N+\overline{N}\rightarrow\chi+\overline{\chi}$ (when the mediator is in equilibrium with
the SM). Also, we have used the relativistic approximations 
$\langle(\sigma v)_{\chi\chi NN}\rangle_{T}^{\rm rel}\simeq\frac{\lambda^{4}}{32\pi g_{\chi}^{2}T^{2}}$, $\langle(\sigma v)_{\phi\phi NN}\rangle_{T}^{\rm rel}\simeq\frac{\lambda^{4}}{32\pi g_{\phi}^{2}T^{2}}[\ln(2T/m_{\chi})-\gamma]$, where $\gamma\approx0.577$ is the Euler-Mascheroni constant,
and the fact that $N$ is in equilibrium with the SM, 
$n_{N}=n_{N}^{\rm eq}(T)$. It is worth emphasizing that the first and second terms in Eq.~(\ref{eq:freeze-in-NN-x}) correspond to the processes $N+\overline N \rightarrow \chi + \overline \chi$ and $N+\overline N \rightarrow \phi + \overline \phi$, respectively.
To get the final relic abundance, we evaluate the first term at 
$x=x_{\rm fin,\chi}\approx c_{2} m_{N}/2m_{\chi}$
and the second at $x=x_{\rm fin, \phi}\approx c_{3} m_{N}/(2m_{\phi})$, leading to 
\begin{equation}
\label{eq:freeze-in-NN-xfin}
Y_{\chi}^{\rm fin}=\frac{\lambda^{4}m_{N}^{5}}{64\pi^{5}H(m_{N})s(m_{N})m_{\chi}}\left\{c_{2}+c_{3}\frac{m_{\chi}}{m_{\phi}}\left[1-\gamma+\ln\left(\frac{4\, m_{\phi}}{c_3 \,m_{\chi}}\right)\right]\right\}.
\end{equation}
For the benchmark $m_{\chi}=2~{\rm TeV}$, $m_{\phi}=3~{\rm TeV}$, we find $c_2=c_3 \approx 1.61$. This expression holds in the blue region of Fig.~\ref{fig:paramsummary}, and since it is independent of $y$, it gives rise to a straight horizontal line for the correct relic abundance in that region. The boundary between the two behaviors (boundary between the blue and green regions in Fig.~\ref{fig:paramsummary}) can be determined by comparing equations (\ref{eq:freeze-in-LH-xfin}) and (\ref{eq:freeze-in-NN-xfin}). In particular, we find
\begin{equation}
y < \lambda\sqrt{\frac{c_{2}}{c_{1}}\frac{m_{\phi}}{m_{\chi}}+\frac{c_{3}}{c_{1}}\left[1-\gamma+\ln\left(\frac{4}{c_{3}}\frac{m_{\phi}}{m_{\chi}}\right)\right]},
\end{equation}
for the second process to be dominant. For the benchmark $m_N=200~{\rm GeV}$, $m_{\chi}= 2~{\rm TeV}$, $m_{\phi}=3~{\rm TeV}$, we obtain $y \lesssim 2.2 \lambda$.  Moreover, If the yield produced by the processes in Eq. \ref{eq:freeze-in-NN-x} reaches the equilibrium value before the processes shut off, one transitions to the freeze-out region, and thus this equation gives the horizontal boundary line between the green and white regions.
In this regime, $\lambda$ is large enough for the $\phi$ (inverse) decay to reach equilibrium before DM freeze-in happens. Increasing $\lambda$ further would result in fully-thermalizing the dark sector with the SM, and DM would be produced by standard freeze-out. Fig.~\ref{fig: standard freeze-in} shows the behavior of the dark sector yields and temperature for a benchmark point in this regime. 

\subsection{Double freeze-in}
\begin{figure}[ht]
\centering
\includegraphics[width=0.49\textwidth]{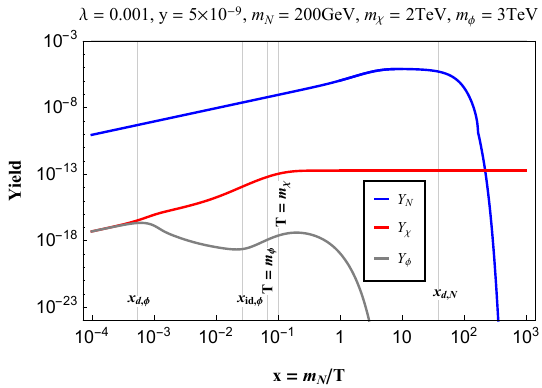}~~
\includegraphics[width=0.49\textwidth]{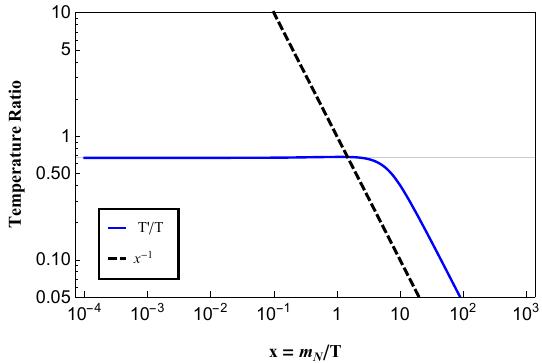}
\caption{The dark sector yields (left panel) and their temperature relative
to the visible temperature (right panel) in the double freeze-in
regime. The coupling $y$ is small so that $N$ (blue line) is not
in equilibrium with the SM bath but is frozen in through the scattering of
bath particles until those processes shut off (around $x\sim1$),
and it later decays (around $x\sim x_{{\rm d},N}$) back to bath particles. The coupling $\lambda$
is large enough such that the process $N+\overline{N}\rightarrow\chi+\overline{\chi}$
freezes in the DM density (red line). The scalar $\phi$ (gray
line) is first produced by the SM bath 
but has a large decay rate so its yield decreases as $x^{-2}$, see the thesis \cite{ElDawBaboAbdelrahim:2023rll}. This continues until its inverse decay becomes appreciable
and increases its density, only to be decreased again by becoming
non-relativistic. The proximity of these two events precludes any
significant production of $\phi$ and hence equilibration. The temperature
ratio remains at its freeze-in value of $2/3$ when $N$ is relativistic
and falls as $x^{-1}$ in the non-relativistic regime. 
}
\label{fig:double freeze-in}
\end{figure}

When $N$ cannot establish equilibrium with the SM (small $y$),
and all dark particles are produced by freeze-in, we dub this production scenario
 double freeze-in. Although we still have the two competing processes
described above, in the standard freeze-in regime, we find that the
correct relic abundance is only obtained when the process $N+\overline{N}\rightarrow\chi+\overline{\chi}$
dominates DM production, and hence the ``double'' freeze-in (${\rm SM}\rightarrow N\rightarrow\chi$).
One still needs to follow the total $\phi+\chi$ yield and Eq.~(\ref{eq:phi+chi yield}) 
applies, however, the number density of $N$ is not the equilibrium
number density, but it is instead given by Eq.~(\ref{eq: early N yield}). Putting
these together and using Eq.~(\ref{eq:early temp}) for the dark temperature,  we rewrite the last two terms in Eq.~(\ref{eq:phi+chi yield}) and integrate to obtain
\begin{equation}
\label{eq: double fin sum}
Y_{\phi+\chi}=Y_{\chi}^{\rm eq, rel}\frac{3x^{3}}{4x_{N\chi}x_{{\rm scat},N}^{2}} + 
Y_{\chi}^{\rm eq, rel}\frac{3x^{3}}{4x_{N\chi}x_{{\rm scat},N}^{2}}
\left[\frac{1}{3}-\gamma+\ln\left(\frac{4\, m_{N}}{3\, x\, m_{\chi}}\right)\right].
\end{equation}
We note that the first and second terms in Eq.~(\ref{eq: double fin sum}) correspond to the processes $N+\overline N \rightarrow \chi + \overline \chi$ and $N+\overline N \rightarrow \phi + \overline \phi$, respectively.
The final DM relic abundance is found by evaluating the first term at  
$x = x_{\rm fin, \chi}\approx c_{2} m_{N}/(3 m_{\chi})$
and the second at $x = x_{\rm fin, \phi}\sim c_{3}m_{N}/(3 m_{\phi})$, where
the factor of 3 (as opposed to 2 in the corresponding expressions
in the standard freeze-in regime) comes from the fact that the dark
temperature (\ref{eq:early temp}) is the relevant temperature here. The result is
\begin{equation}
Y_{\chi}^{\rm fin}=\frac{9y_{t}^{4}\lambda^{4}y^{4}m_{N}^{9}}{2^{19}\pi^{11}g_{N}^{2}H(m_{N})^{3}s(m_{N})m_{\chi}^{3}}
\left\{c_{2}^{3}+c_{3}^{3}\, \frac{m_{\chi}^{3}}{m_{\phi}^{3}}
\left[\frac{1}{3}-\gamma+\ln\left(\frac{4 \, m_{\phi}}{c_3 \, m_{\chi}}\right)\right]\right\}.
\end{equation}
For the benchmark $m_{\chi}=2~{\rm TeV}$, $m_{\phi}=3~{\rm TeV}$, we find $c_2=c_3 \approx 2.59$. 
If we compare this to the relic abundance produced by the process
$L+\overline{H}\rightarrow\chi+\overline{\phi}$, we get a boundary in the $y-\lambda$ plane
separating the two behaviors (boundary between the brown and pink regions in Fig.~\ref{fig:paramsummary}): 
\begin{equation}
\lambda \, y <\frac{64 \, \pi^{3} \, g_{N}  \, H(m_N) \, m_\chi}{3 \, y_{t}^{2} \, m_N^2}  
\sqrt{\frac{2\, c_1 \, m_\chi }{m_\phi}}
\left\{c_{2}^{3}+c_{3}^{3}\, \frac{m_{\chi}^{3}}{m_{\phi}^{3}}
\left[\frac{1}{3}-\gamma+\ln\left(\frac{4 \, m_{\phi}}{c_3 \, m_{\chi}}\right)\right]\right\}^{-1/2}.
\end{equation}
For the benchmark $m_N=200~{\rm GeV}$, $m_{\chi}=2~{\rm TeV}$, $m_{\phi}=3~{\rm TeV}$, we obtain $\lambda \,y \lesssim 9.8 \times 10^{-13}$.
Fig.~\ref{fig:double freeze-in} shows the solution to the Boltzmann equations in this regime.

\subsection{Dark sector equilibration}
\begin{figure}
\centering
\includegraphics[width=0.47\textwidth]{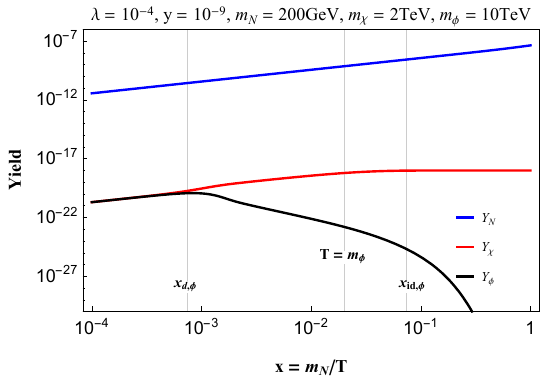}~~
\includegraphics[width=0.47\textwidth]{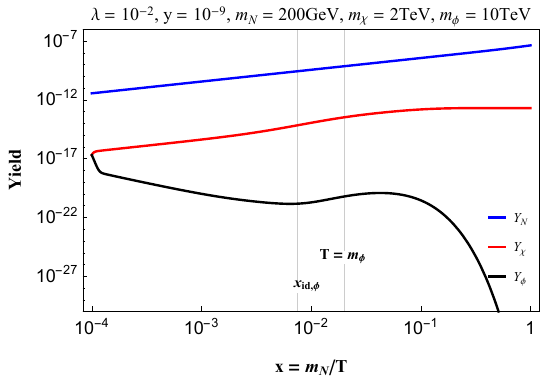}\\
\includegraphics[width=0.47\textwidth]{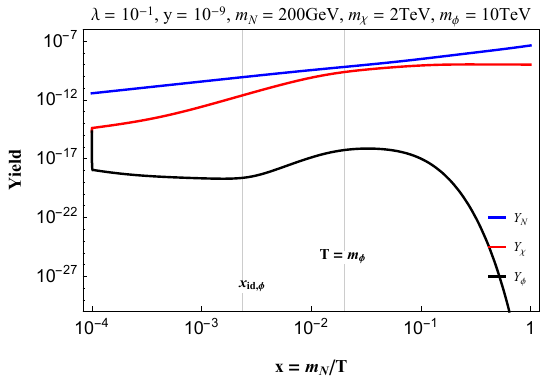}~~
\includegraphics[width=0.47\textwidth]{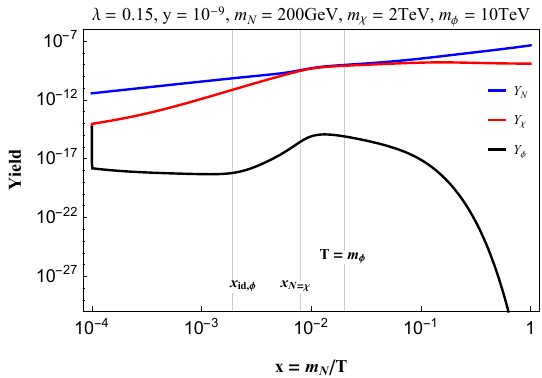}\\
\includegraphics[width=0.47\textwidth]{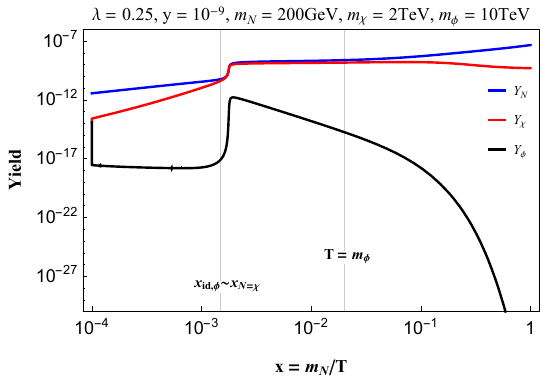}~~
\includegraphics[width=0.47\textwidth]{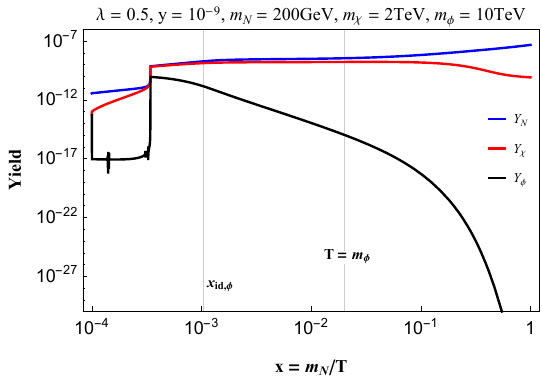}
\caption{{\it Dark sector approach to equilibrium: Yields.}  The successive frames show the dark yields as the coupling $\lambda$ is increased from the freeze-in regime until equilibrium is established. The $\phi$ bottleneck: $\phi$ decay always causes its density to be smaller than $\chi$ and $N$. Its density can only be increased via the process $\bar{N}+\chi\rightarrow\phi$, which is suppressed until the $\chi$ density becomes large early enough, 
by the time-scale $x_{{\rm id},\phi}<m_{\phi}/m_{N}$. The process becomes as efficient as possible when $\chi$ and $N$ densities become comparable and establish chemical equilibrium early enough, resulting in explosive production of $\phi$ particles. Note that the $\chi$ density varies as $\lambda^4$, so that the time at which the $\chi$ and $N$ densities become comparable is more sensitive than the time-scale $x_{{\rm id},\phi}$ to increases in $\lambda$.}
\label{fig: 15 R3a}
\end{figure}

\begin{figure}
\centering
\includegraphics[width=0.47\textwidth]{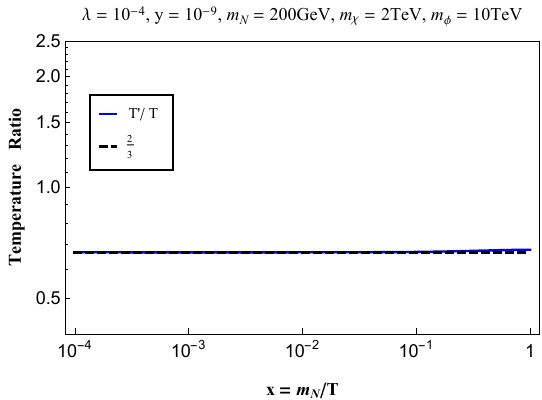}~~
\includegraphics[width=0.47\textwidth]{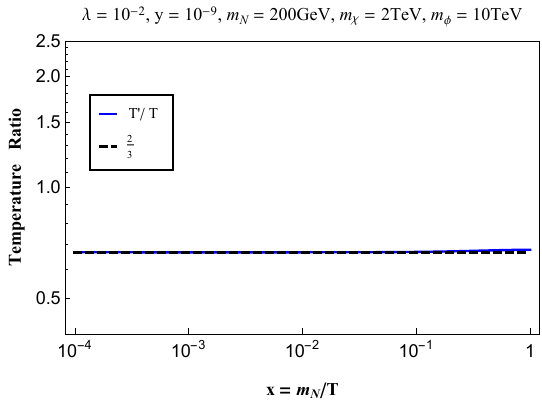}\\
\includegraphics[width=0.47\textwidth]{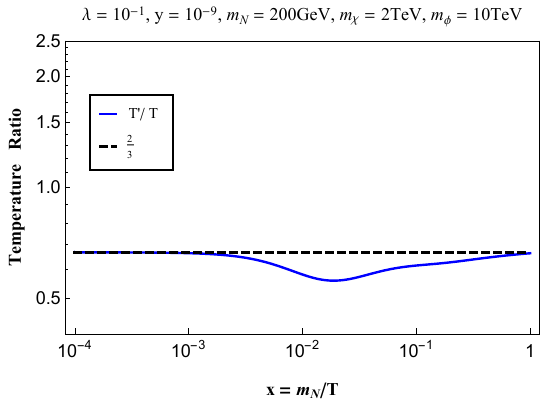}~~
\includegraphics[width=0.47\textwidth]{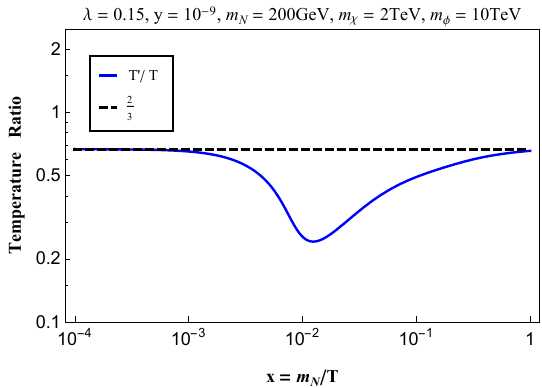}\\
\includegraphics[width=0.47\textwidth]{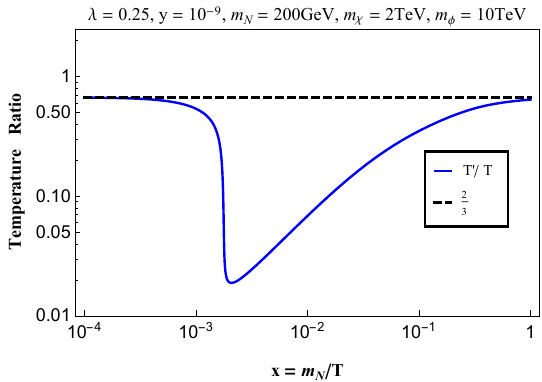}~~
\includegraphics[width=0.47\textwidth]{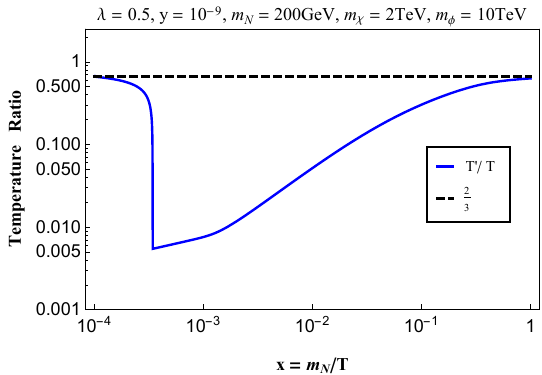}
\caption{{\it Dark sector approach to equilibrium: Temperature.}  The successive frames show the ratio of the dark to visible temperature as the coupling $\lambda$ is increased from the freeze-in region. The ratio is 2/3 in the freeze-in region, Eq.~(\ref{eq:early temp}), when dark particles have feeble interactions. As $\lambda$ is increased and dark particle production becomes efficient, the energy leaked from the SM gets shared among a larger and larger number of particles, driving their temperature to decrease. This decrease stops when particle production ceases due to particles becoming non-relativistic or after full equilibrium is reached. The subsequent increase in the temperature ratio is due to SM production of $N$ still being efficient. The thermal equilibrium period can be seen in the last frame.}
\label{fig: 16 R3b}
\end{figure}

In the regime where the dark sector has a very weak coupling to the
SM, so that its temperature is different from that of the bath,
a large dark interaction can lead to the formation of a distinct thermal
bath in the dark sector with a temperature much smaller than the visible
sector temperature. Recall that in the freeze-in regime, the dark
particles are produced with a temperature of the order of the visible
sector temperature, $T^\prime =\frac{2}{3}T$, but their number density is
small (i.e., there is a large negative chemical potential) compared to the equilibrium
number density. In order to reach full thermal equilibrium, where
the number density is determined by the dark temperature only and the chemical
potential vanishes, one needs to increase the number density and/or
decrease the temperature. Let's understand this thermalization process
in detail: 
\begin{itemize}
\item First, one needs all chemical potentials to vanish. The $2\rightarrow2$
dark processes $N+\overline{N}\leftrightarrow\chi+\overline{\chi},\phi+\overline{\phi}$
and $\phi+\overline{\phi}\leftrightarrow\chi+\overline{\chi}$ , when in chemical
equilibrium, can only set the chemical potentials equal to each other,
$\mu_{N}=\mu_{\chi}=\mu_{\phi}$, but cannot guarantee that they equal
zero. The process crucial to establish this is the (inverse) decay
process $\phi\leftrightarrow\chi+\overline{N}$, which is the only dark
process that changes total dark particle number.
\item The total dark number density governs the behavior of the dark temperature
as is shown by Eq.~(\ref{eq:TPdivT-1}) 
for the temperature ratio; increasing the
dark number density decreases the temperature because their product gives
the fixed energy leaked to the dark sector from the visible one. The
total dark yield is given by the equation:
\begin{eqnarray}
Hsx\frac{dY_{\rm tot}}{dx} &=& \langle\Gamma_{N,{\rm tot}}\rangle_{T} \, n_{N}^{\rm eq}(T)
+2 \, \langle(\sigma v)_{\chi\phi LH}\rangle_{T} \, n_{\chi}^{\rm eq}(T) \, n_{\phi}^{\rm eq}(T)\nonumber \\ & &
+\langle\Gamma_{\phi}\rangle_{T'}\left(n_{\phi}-n_{\phi}^{\rm eq}(T')\frac{n_{\chi}}{n_{\chi}^{\rm eq}(T')}\frac{n_{N}}{n_{N}^{\rm eq}(T')}\right),
\end{eqnarray}
and its dependence on the dark sector comes through the last term only.
\item Full thermal equilibrium means a similar abundance for all the particles.
Let's start in the freeze-in regime: $N$ is most abundant, followed
by $\chi$ whose abundance grows with increasing dark interaction
$\lambda$. However, the $\phi$ density is quite small since its
decay rate is large and its production is quenched by the small density
of the other particles. Therefore, the transition from the freeze-in
regime to a thermal dark bath has to start with overcoming this $\phi$ 
bottleneck. Notice that the production processes $L+\overline{H}\rightarrow\chi+\overline{\phi}$
and $N+\overline{N}\rightarrow\phi+\overline{\phi}$ , when the decay is active,
result in the density of $\phi$ decreasing as a power law instead
of the familiar exponential (see the thesis \cite{ElDawBaboAbdelrahim:2023rll}), but they cannot increase the $\phi$ density. 
\item The first step to increase the $\phi$ density is to turn-on
the inverse decay process by having $\chi$ be abundant enough to
render its scattering off of $N$ efficient, thus allowing chemical equilibrium
of the decay/inverse decay. Clearly, this becomes as efficient as
possible when $\chi$ and $N$ are the most abundant, and hence the
next step is the process $\chi+\overline{\chi}\leftrightarrow N+\overline{N}$ establishing
chemical equilibrium. This results in the explosive creation of $\phi$ particles where the yield of $\phi$ increases by many orders of magnitude
almost instantaneously.  This condition of simultaneous equilibrium of the (inverse) decay of $\phi$ and $\chi-N$ annihilation determines the boundary between this region and the double freeze-in region. For quantitative details, see \cite{ElDawBaboAbdelrahim:2023rll}.
\item These two processes lead to $\mu_{\phi}=2\mu_{\chi}=2\mu_{N}$, and
the final step to complete thermalization is for one of the two processes:
$\phi+\overline{\phi}\rightarrow N+\overline{N}$ or $\phi+\overline{\phi}\rightarrow\chi+\overline{\chi}$
to enter chemical equilibrium.  Figs.~\ref{fig: 15 R3a} and \ref{fig: 16 R3b} show the behavior of the dark yields and temperature as the dark interaction is dialed up and the dark sector approaches equilibrium.
\end{itemize} 

Once a thermal bath at temperature $T^{\prime}$ is established, the
details of how the thermalization came about become unimportant and do not affect the
subsequent evolution of the system. If $T^{\prime}>m_{\phi}$, all
dark particles are relativistic and their yields are
\begin{equation}\label{eq: 32 thermal yields}
Y_{i}=\frac{g_{i}}{\pi^{2}}\frac{m_{N}^{3}}{s(m_{N})}\frac{T'^{3}}{T^{3}}=Y_{i}^{\rm eq, rel}\frac{T'^3}{T^3},
\end{equation}
where $i=N,\chi,\phi$. The temperature ratio can be obtained from
Eq.~(\ref{eq:TPdivT-1}) 
by using the yields above, 
\begin{equation}
\label{eq: Temp ratio}
\frac{T'}{T}=\left[\frac{2}{3}\frac{1}{g_{N}+g_{\chi}+g_{\phi}}\left(g_N\frac{x}{x_{{\rm scat},N}}+2g_{\chi}\frac{x}{x_{LH}}\right)\right]^{1/4}, 
\end{equation}
which grows with time, i.e.,  the dark sector is heated relative to the visible
sector. However, the dark temperature itself decreases with time, $T^{\prime}\propto x^{-3/4}$,
which is
slower than the familiar $x^{-1}$ expected for relativistic species
in an expanding universe. The relative heating is due to the energetic
particles injected from the SM via the connecting processes in Fig.~\ref{fig:connect-diagrams}. 

The next
important event in the dark history marks the end of the period of
thermalization. When the dark temperature falls below the mass of
$\phi$ (the heaviest dark particle), its density becomes Boltzmann-suppressed and it decouples from the thermal bath at a time $x_{\rm{dec},\phi}$
determined by the condition 
\begin{equation}
\label{eq:phi-decouple}
n_{\phi}^{\rm eq}(T^{\prime})\langle(\sigma v)_{\phi,{\rm tot}}\rangle^{\rm Nrel}=H,
\end{equation}
where $\langle(\sigma v)_{\phi,\rm{tot}}\rangle^{\rm Nrel}=\langle(\sigma v)_{\phi\phi NN}\rangle^{\rm Nrel}+\langle(\sigma v)_{\phi\phi\chi\chi}\rangle^{\rm Nrel}$ is the total thermally-averaged cross section in the non-relativistic regime (see Appendix~\ref{app:therm-avg}). This leads to 
\begin{equation}
x_{\rm{dec},\phi}= \left(\frac{m_N}{m_{\phi}}\left(\frac{2}{3}\frac{g_N}{g_N+g_{\chi}}\frac{1}{x_{{\rm scat},N}}\right)^{1/4}\ln\left[\frac{g_{\phi}}{(2\pi)^{3/2}}m_{\phi}^{1/3}m_N^{8/3}\frac{\langle(\sigma v)_{\phi,{\rm tot}}\rangle}{H(m_N)}\left(\frac{2}{3}\frac{g_N}{g_N+g_{\chi}}\frac{1}{x_{{\rm scat},N}}\right)^{2/3}\right]\right)^{4/3}.   
\end{equation}

\begin{figure}
\includegraphics[width=0.47\textwidth]{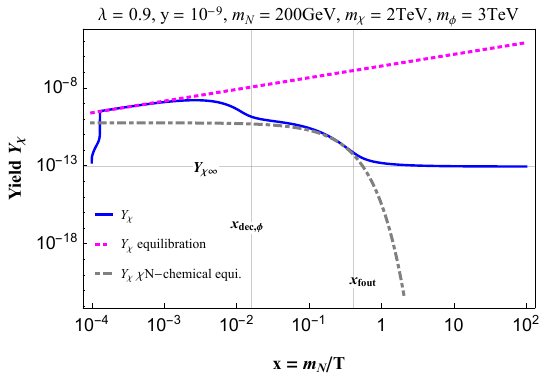}~~\includegraphics[width=0.45\textwidth]{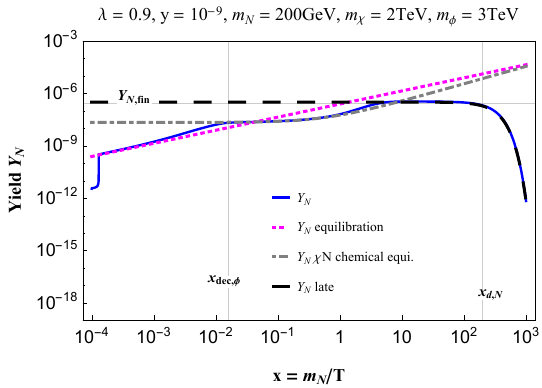} \\
%
\includegraphics[width=0.47\textwidth]{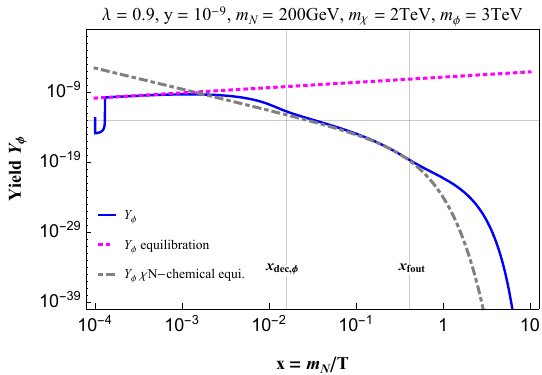}~~\includegraphics[width=0.45\textwidth]{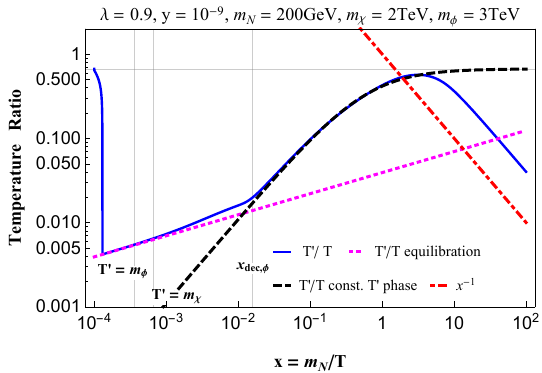}
\caption{{\it DM dark freeze-out.} The dark sector establishes thermal equilibrium at temperature $T'$ by increasing the dark number density and decreasing the temperature as can be seen in the very first part of all the plots. Once in thermal equilibrium, the temperature ratio and dark yields increase slowly (Eqs.~(\ref{eq: Temp ratio},\ref{eq: 32 thermal yields})) due to particle production from the SM as shown by the red dotted lines. As the dark temperature falls below the $\phi$ mass, its yield decreases as it annihilates. As a result of the strong interaction between $\chi$ and $\phi$ and their closeness in mass, $\chi$ co-annihilates with $\phi$ until $\phi$ decouples at $x_{\rm dec,\phi}$. Thermal equilibrium is lost when $\phi$ decouples, but $\phi$ decay remains in chemical equilibrium as shown by the dot-dashed line (Eq.~(\ref{eq: 245 phi yield})). $\chi$ and $N$ also maintain chemical equilibrium as shown by the dot-dashed lines (Eq.~(\ref{eq: chi R3 che eq}) and the similar one for $N$). In this regime, the relative heating of the dark sector is more efficient, since the particles produced from the SM do not thermalize; the dark temperature is kept constant. As the production of $N$ ceases, $N$ freezes in, DM annihilates and freeze-out at $x_{\rm fout}$ to the value (\ref{eq: freeze-out yield}). The dark sector is now fully non-relativistic and its temperature scales as $x^{-2}$. Note that the $\phi$ yield follows that of $\chi$ and $N$, because the decay remains in chemical equilibrium.}
\label{fig: 18 R3c}
\end{figure}

With the two processes $\phi+\overline{\phi}\rightarrow\chi+\overline{\chi},N+\overline{N}$
falling out of chemical equilibrium, full thermal equilibrium is lost
in the dark sector as one needs at least three processes for the chemical
potentials to vanish. The other two processes $\phi\leftrightarrow\chi+\overline{N}$,
$\chi+\overline{\chi}\leftrightarrow N+\overline{N}$ remain in chemical equilibrium.
Together with dark particle
production from the SM, they represent the
important processes determining dark sector evolution at this time. For $\phi$, the process $\phi\leftrightarrow\chi+\overline{N}$ keeps it in chemical equilibrium with $\chi$ and $N$. Its yield is
\begin{equation}\label{eq: 245 phi yield}
    Y_{\phi}=Y_{\phi}^{\rm eq}(T') \frac{Y_{N}}{Y_{N}^{\rm eq}(T')}\frac{Y_{\chi}}{Y_{\chi}^{\rm eq}(T')},
\end{equation}
and hence it is enough to focus on the $\chi-N$ system. The total $\chi+N$ yield
obeys
\begin{equation}
H s x \frac{dY_{\chi+N}}{dx}= \langle\Gamma_{N,{\rm tot}}\rangle_{T} \, n_{N}^{\rm eq}(T)
+2\langle(\sigma v)_{\chi\phi LH}\rangle_{T} \, n_{\chi}^{\rm eq}(T) \, n_{\phi}^{\rm eq}(T),
\end{equation}
where the process $L+\overline{H}\rightarrow\chi+\overline{\phi}$ is included because $\phi$ might still be relativistic with respect to the SM temperature, and the factor of 2 is due to the eventual dumping of the $\phi$ density into $\chi$ and $N$. . The solution when this is the case is given by
\begin{equation}\label{eq: R3 total Nchi yield}
Y_{\chi+N}=Y_{\chi+N}(x_{\rm{dec},\phi})+Y_{N}^{\rm eq,rel}\frac{x-x_{\rm{dec},\phi}}{x_{{\rm scat},N}}+2 Y_{\chi}^{\rm eq,rel}\frac{x-x_{\rm{dec},\phi}}{x_{LH}},
\end{equation}
where $Y_{\chi+N}(x_{\rm{dec},\phi})$ is the sum of $\chi$ and $N$ equilibrium yields
at the end of the thermal phase. 
Since $\chi$ and $N$ are in chemical
equilibrium, their proportions are fixed and this allows determination
of the individual yields,
\begin{equation}\label{eq: chi R3 che eq}
Y_{\chi}=\frac{Y_{\chi}^{\rm eq}(T')}{Y_{\chi}^{\rm eq}(T')+Y_{N}^{\rm eq}(T')}Y_{\chi+N}.
\end{equation}
As for the dark temperature, Eqs.~(\ref{eq:TPdivT-1},\ref{eq: R3 total Nchi yield}) 
imply 
\begin{equation}
\label{eq: Temp 2phases}
\frac{T'}{T}=\frac{2}{3}\frac{Y_{N}^{\rm eq,rel} \frac{x}{x_{{\rm scat},N}}+2\,Y_{\chi}^{\rm eq,rel}\frac{x}{x_{LH}}}{Y_{\chi+N}(x_{{\rm dec},\phi})+Y_{N}^{\rm eq,rel}\frac{x}{x_{{\rm scat},N}}+ 2 \, Y_{\chi}^{\rm eq,rel}\frac{x}{x_{LH}}}.
\end{equation}
From here we see there are two phases in the temperature evolution. 
The first phase corresponds to the first term in the denominator
dominating leading to the dark temperature being constant in time. This results in the $\chi$ yield, Eq.~(\ref{eq: chi R3 che eq}), being constant.
Although $\chi$
might be non-relativistic at this point, the active injection of energetic
$N$s (that cannot thermalize but are immediately converted into $\chi$
via the still-efficient $N+\overline{N}\rightarrow\chi+\overline{\chi}$ keeping
the total number constant) act to counter the Boltzmann suppression
and keep the DM yield intact. As the universe expands, the second
term in the denominator of Eq.~(\ref{eq: Temp 2phases}) 
starts to become important,
driving the temperature ratio to its freeze-in value of 2/3. This phase is interrupted by the SM temperature falling below the
mass of $N$, signaling the end of energy injection from the SM. The
dark sector is then fully non-relativistic and decoupled, and hence its
temperature decreases like $a^{-2}$. This behavior is shown in Fig.~\ref{fig: 18 R3c}.

\begin{figure}
\includegraphics[width=0.45\textwidth]{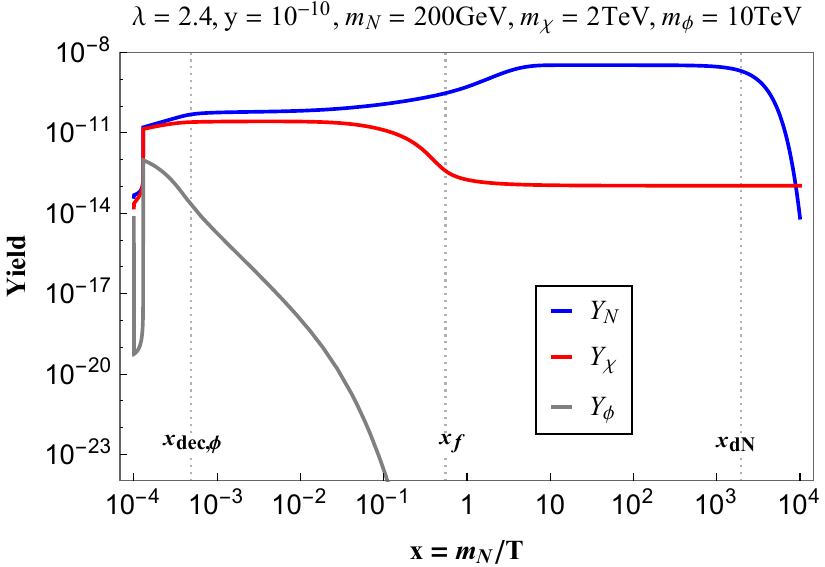}~~\includegraphics[width=0.45\textwidth]{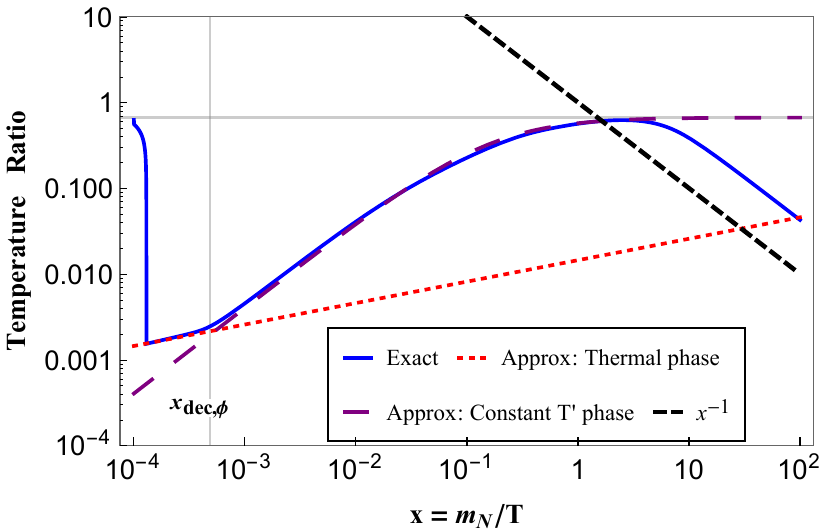}
\caption{{\it DM dark freeze-out: increasing $m_{\phi}$}. Compared to Fig.~\ref{fig: 18 R3c}, for which the mass of $\phi$ is smaller, the equilibration region shrinks. The region of constant dark temperature enlarges, and the co-annihilation feature in it disappears. The $\chi$ yield remains constant in this region, even though $\chi$ is non-relativistic. This is due to the SM production of $N$s and their efficient conversion into $\chi$s, since $\chi$ and $N$ are in chemical equilibrium. If $m_{\phi}$ is increased further, the thermal phase is lost.}.  
\label{fig: 19 R3c}
\end{figure}

The freeze-out yield of DM is determined in the usual way and is
given by: 
\begin{equation}\label{eq: freeze-out yield}
Y_{\chi}^{\rm fout}=\sqrt{\frac{45}{\pi g_{*}}}\frac{x_{\rm fout}}{\langle(\sigma v)_{\chi\chi NN}\rangle^{\rm Nrel}m_{N}M_{P}},
\end{equation}
where the freeze-out time $x_{\rm fout}$ is determined by $n_{\chi} \langle(\sigma v)_{\chi\chi NN}\rangle^{\rm Nrel} =H$, where one uses 
Eqs.~(\ref{eq: R3 total Nchi yield}-\ref{eq: Temp 2phases}) 
to get $n_{\chi}$.  

\subsection{Standard freeze-out} 
\begin{figure}
\includegraphics[width=0.7\textwidth]{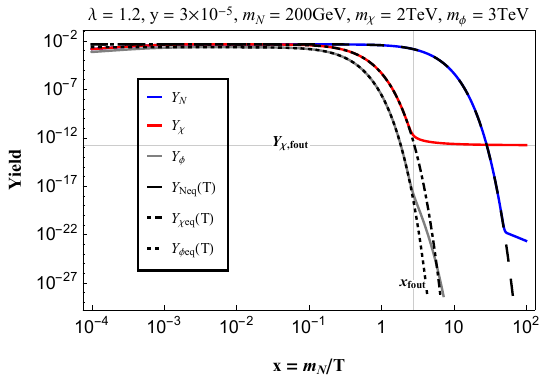}
\caption{{\it DM freeze-out: WIMP case.} All dark particles establish equilibrium with the SM bath, as shown by the dashed lines of equilibrium at temperature $T$. At late times, DM freezes out, while the other particles decay in equilibrium. $\phi$ decay maintains chemical equilibrium and the $\phi$ yield is determined by those of DM and mediator (hence the feature in the $\phi$ yield as DM freezes out). There are late annihilations of DM into $N$s as can be seen in the lower right.}  
\label{fig: WIMP}
\end{figure}

In the large $y$, large $\lambda$ corner of parameter space,
the dark sector can reach kinetic and chemical equilibrium with the
SM bath before DM freezes out, i.e. $T'=T$, $Y_{N}=Y_{N}^{\rm eq}(T)$,
$Y_{\phi}=Y_{\phi}^{\rm eq}(T)$, $Y_{\chi}=Y_{\chi}^{\rm eq}(T)$. Considering
our mass hierarchy (Eq.~(\ref{eq:mass-hierarchy})), one sees that DM freezes out of this equilibrium by annihilation $\chi+\overline{\chi}\rightarrow N+\overline{N}$.
The DM yield after freeze-out is thus given by Eq.~(\ref{eq: freeze-out yield}),
but with $x_{\rm fout}\simeq\ln\left(0.04 c (c+2) \frac{g_{\chi}}{\sqrt{g_{*}}}\langle(\sigma v)_{\chi\chi NN}\rangle^{\rm Nrel} m_{\chi}M_{P}\right)$ which is the usual WIMP one. Note that
this depends on $\lambda$ only and is independent of $y$. Fig.~(\ref{fig: WIMP})
shows the freeze-out of DM in this case.

It is worth mentioning that our results are independent of the initial condition, i.e., the initial temperature of the visible sector (as long as it is chosen to be above all the relevant masses). The insensitivity of the thermal histories to the initial condition can be attributed to the beginning of populating the dark sector (starting from zero particles in the dark sector) which is basically a freeze-in process. Since our interactions are described by renormalizable operators, the freeze-in is always IR-dominated (independent of the maximum temperature or in our case the initial temperature)~\cite{Hall:2009bx}, in contrast to the UV-dominated freeze-in which occurs through non-renormalizable operators and depends on the highest temperature~\cite{Elahi:2014fsa}.

\section{Signals}
\label{sec:signals}

\subsection{Indirect detection}
The possible annihilation of DM at present time in astrophysical environments such as the center of the Milky Way and dwarf galaxies which have high concentrations of DM can lead to the production of gamma rays in the final state through hadronization, radiation, and decay of the SM particles. 
The annihilations of a DM candidate with mass $m_\text{DM}$ that is not its own antiparticle, in a solid angle $\Delta\Omega$ result in gamma rays with differential energy flux given by
\begin{equation}
    \frac{d\Phi_\gamma}{d E_\gamma}(E_\gamma,\Delta\Omega)=\frac{\langle\sigma v\rangle}{16\pi m_\text{DM}^2}\sum_f {\rm Br}_f\frac{dN_\gamma^f}{dE_\gamma}(E_\gamma) J(\Delta\Omega),
\end{equation}
with 
\begin{equation}
    J(\Delta\Omega)=\int_{\Delta\Omega}\int_\text{los}\rho^2[s(r,\theta)] \, ds  \, d\Omega,
\end{equation}
where $\langle\sigma v\rangle$ is the thermally averaged annihilation cross section, $dN_\gamma^f(E_\gamma)/dE_\gamma$ is 
the number of photons at a given energy $E_\gamma$ produced in the annihilation channel with final state $f$ with branching ratio ${\rm Br}_f$.
$J(\Delta\Omega)$, the so-called $J$-factor, which contains all the astrophysical information, is evaluated by integrating over DM local density, $\rho(s)$, along the line of sight (los), $s(r,\theta)$.    

No significant gamma-ray excess at very high energies ($\gtrsim 100\,\text{GeV}$) in the Inner Galaxy Survey of the High Energy Stereoscopic System
(H.E.S.S.)~\cite{HESS:2022ygk} leads to stringent and robust constraints to DM models. The statistical analysis used to convert H.E.S.S. data into limits on DM annihilation is based on a two dimensional log-likelihood ratio test statistic. In this analysis, the expected spatial and spectral features of the signal over energy bins and spatial bins in regions of interest are used to build the  likelihood function~\cite{HESS:2022ygk}.

Rather than perform a log-likelihood analysis, we will employ a simple procedure to translate the already existing constraints on DM annihilations to pairs of $W$ bosons to the neutrino portal model. 
Since the primary decay channel of weak-scale $N$s is $N\rightarrow Wl$, this procedure should yield a reasonably accurate bound on the $\chi \overline \chi \rightarrow N \overline N$ annihilation cross section. DM annihilation into $N$s, $\chi+\overline{\chi}\rightarrow N+\overline N$, followed by decay of $N$ to $W$s, $N\rightarrow Wl$ results in a distribution of $W$s over the kinematically accessible energies.
If the spectrum of emitted $W$s in $N$s rest frame with energy $E'$ is $dN'_W/dE'_W$, the spectrum of these particles boosted to the rest frame of $\chi$ is given by $dN_W/dE_W$ as~\cite{Batell:2017rol}
\begin{eqnarray}
\frac{dN_W}{dE_W} =  \int_{\gamma(E_W-\beta\sqrt{E_W^2-m_W^2})} ^{\gamma(E_W+\beta\sqrt{E_W^2-m_W^2})} \frac{dE_W'}{2\beta\gamma \sqrt{E_W'^2-m_W^2}}\,\frac{dN'_W}{dE'_W}, ~~~~~ \gamma = (1-\beta^2)^{-1/2} = m_\chi/m_N. ~~~
\label{eq:spectrum-boosted}
\end{eqnarray}
In the rest frame of $N$, $W$ is monoenergetic,
\begin{equation}
\frac{dN'_W}{dE'_W} =2\delta\left[E'_W-\left(\frac{m_N^2+m_W^2}{2m_N}\right)\right].
\end{equation}
This monoenergetic distribution gives rise to a box-like boosted distribution in the rest frame of $\chi$ given by
\begin{equation}
\frac{dN_W}{dE_W}=\frac{2m_N^2}{(m_N^2-m_W^2)\sqrt{m_\chi^2-m_N^2}}\Theta(E_W-E_-)\Theta(E_+-E_W),
\end{equation}
where $\Theta$ is the Heaviside step function and $E_{\pm}=\frac{m_\chi}{2}\left[1+\frac{m_W^2}{m_N^2}\pm\left(1-\frac{m_W^2}{m_N^2}\right)\sqrt{1-\frac{m_N^2}{m_\chi^2}}\right]$.

The constraints on $W$ channel, $\langle\sigma v\rangle_W$, weighted by the energy distribution of $W$'s in the rest frame of $\chi$, $dN_W/dE_W$, can be used to estimate the constraints on $N$ channel, $\langle\sigma v\rangle_N$, as:
\begin{equation}
\langle\sigma v\rangle_N(m_\chi,m_N)=\frac{2m_N^2}{(m_N^2-m_W^2)\sqrt{m_\chi^2-m_N^2}}
\int_{E_- }^{E_+}\langle\sigma v\rangle_W(E_W) dE_W.
\label{eq:sigmavweightedlimit}
\end{equation}
No significant gamma-ray excess is observed at very high energies ($\gtrsim 100\,\text{GeV}$) in the Inner Galaxy Survey of the High Energy Stereoscopic System
(H.E.S.S.)~\cite{HESS:2022ygk}, which leads to an upper bound on the coupling $\lambda$. 
For $m_\phi=3\,\text{TeV}$, $m_\chi=2\,\text{TeV}$, and $m_N=200\,\text{GeV}$, and an Einasto profile for the DM halo,  H.E.S.S. searches for gamma-rays from DM annihilation into $W^+W^-$ constrain 
the coupling constant in the dark sector to be 
$\lambda\lesssim 1.9$. A naive estimate of the coupling constant based on $\langle\sigma v\rangle_N(m_\chi,m_N)= \langle\sigma v\rangle_W(m_\chi)$ (instead of Eq.~(\ref{eq:sigmavweightedlimit})) leads to a stronger constraint of $\lambda\lesssim 1.5$. The projected sensitivity of the Cherenkov Telescope Array (CTA)~\cite{CTA:2020qlo} as the world’s most sensitive gamma-ray telescope will be able to reach the canonical thermal relic cross-section for TeV-mass DM. For the neutrino portal model, CTA will be able to probe couplings down to $\lambda\simeq 1.3$ for benchmark values  of $m_\phi=3\,\text{TeV}$, $m_\chi=2\,\text{TeV}$, and $m_N=200\,\text{GeV}$. 
The superficial choice of $\langle\sigma v\rangle_N(m_\chi,m_N)= \langle\sigma v\rangle_W(m_\chi)$ results in the stronger projection of $\lambda\simeq 1.1$.

The non-relativistic limit of thermally averaged annihilation cross section of DM into sterile neutrino, $\chi+\overline{\chi}\rightarrow N+\overline{N}$, is given by:
\begin{equation}
    \langle\sigma v\rangle_N=\frac{1}{g^2_\chi}\frac{\lambda^4}{2\pi}\frac{\left(m_\chi+m_N\right)^2}{\left(m_\phi^2+m_\chi^2-m_N^2\right)^2}\left(1-\frac{m_N^2}{m_\chi^2}\right)^{1/2}.
\end{equation}

\subsection{Collider signals}
For large values of the coupling $y$ the heavy neutrino $N$ may be produced and searched for at high energy colliders, and there are already a variety of bounds from existing searches. Some of these searches rely on lepton number violating signals that are not predicted in our model, e.g., the analyses of Refs.~\cite{ATLAS:2015gtp,ATLAS:2019kpx,CMS:2012wqj,CMS:2015qur,CMS:2016aro,CMS:2018jxx} and the review \cite{Cai:2017mow}. Perhaps the most relevant and stringent existing collider bound comes from a CMS trilepton searches~\cite{CMS:2018iaf,CMS:2024xdq}, which can arise from $N$ produced via the Drell-Yan process, $q \overline q' \rightarrow W \rightarrow N \ell$, followed by the decay $N\rightarrow \ell \ell \nu$. For $N$ that dominantly mixes with electron-flavor neutrinos, this search places a bound on the mixing angle $|U_e|^2 \lesssim 0.004\,(0.05)$ for $m_N$ = 200 GeV~(500 GeV). For $|U_e|^2 \simeq y^2 v^2/m_N^2$, this translates into a bound on the coupling  $y\lesssim 0.07\,(0.6)$ for $m_N$ = 200 GeV~(500 GeV). 

The HL-LHC and future colliders will be able to significantly improve the reach on the coupling $y$ for heavy neutrinos in the 100 GeV - TeV mass range. For example, a recent study has considered the reach of the HL-LHC and a future 100 TeV hadron collider to heavy Dirac neutrinos, also in the trilepton channel~\cite{Feng:2021eke}. For the HL-LHC, with $\sqrt{s} = 14$ TeV and an integrated luminosity of 3 ab$^{-1}$, a projected 95$\%$ C.L. sensitivity of $|U_e|^2 \lesssim 10^{-4} \,(10^{-3})$ for $m_N$ = 200 GeV (500 GeV) is obtained, which implies a sensitivity to 
 the coupling  $y\lesssim 0.01\,(0.1)$ for $m_N$ = 200 GeV (500 GeV). A future 100 TeV hadron collider can improve on this by a factor of 5-10 in the $|U_e|^2$ depending on the mass. 
 
We note that similar bounds and future sensitivities can be derived for $N$ that dominantly mixes with muon-flavored neutrinos; see, e.g., Ref.~\cite{Feng:2021eke}. 
It would also be very interesting to study the prospects for dominantly $\tau$-flavor mixing at the LHC and future colliders. 
 
\subsection{Precision measurements}

Large value of the coupling $y$ will induce sizable heavy-light neutrino mixing and in turn lead to the non-unitarity of PMNS mixing matrix. This in turn can impact a variety of  electroweak precision observables, including $W$ boson mass, the $W$ and $Z$ boson decays, the weak mixing angle, tests of lepton universality tests, tests of CKM unitarity. Bounds of  $|U_{e}|^2 < 2.2 \times 10^{-3}$ and $|U_\mu|^2 < 9.0 \times 10^{-4}$ have been derived from global studies of the electroweak precision data~\cite{delAguila:2008pw,Akhmedov:2013hec,Antusch:2014woa,Blennow:2016jkn}. 
The corresponding bounds on the Yukawa coupling are $y\lesssim 0.01 - 0.1$ for $N$ masses near the weak scale.

\subsection{Direct detection}
The leading contribution to a direct detection signal comes from the effective $Z$-$\chi$-$\chi$ coupling induced at one-loop~\cite{Batell:2017cmf}. In the limit $m_N\ll m_\chi\ll m_\phi$ the effective operator is
\begin{equation}
{\cal L}_{Z\chi\chi}\simeq \frac{y^2\lambda^2v^2}{64\pi^2(m_\phi^2-m_\chi^2)}\left(1-\frac{\lambda^2v^2}{m_N^2}\right)\frac{g}{\cos\theta_W}Z^\mu\overline\chi \gamma_\mu \left(\frac{1+\gamma^5}{2}\right)  \chi,
\end{equation}
For $m_\chi\gg {\rm GeV}$, this leads to a spin-independent scattering cross section on a nucleon in a Xenon target of
\begin{equation}
\sigma_{\rm SI,Xe}\simeq6\times10^{-48}~{\rm cm}^2\,y^4\lambda^4\left[1-\left(\frac{\lambda v}{m_N}\right)^2\right]^2\left(\frac{\rm TeV^2}{m_\phi^2-m_\chi^2}\right)^2 ,
\end{equation}
This is well below current bounds on TeV-scale DM~\cite{LZ:2024zvo} and for $y,\lambda\lesssim0.1$ largely lies below the ``neutrino floor''~\cite{Gaspert:2021gyj} where DM direct detection is extremely challenging.

\subsection{Late $N$ decay}
If $N$ is long lived, $y\lesssim 4 \times 10^{-7}  \, x^{-1/2}_{\rm fin} \, (m_{N}/200~\rm{ GeV})^{1/2}$ (see Eq.~(\ref{eq:N equili condition}), then it freezes in after becoming non-relativistic, and its energy density
scales as $x^{-3}$ before it decays. If it were to dominate the energy
density of the universe before decaying this would dilute the DM relic abundance. 
We now show that this does not happen in this model in the parameter regions of interest. 
First, note that late time freeze-in energy density of $N$ (before decay) is 
\begin{align}
\rho_{N}=m_{N}n_{N}=\frac{c^3 \, g_{N}  \, m_N^4 }{3\, \pi^{2} \, x_{{\rm id},N}^{3} \, x^3 },
\end{align}
where $n_{N}$ is the freeze-in value corresponding to the yield given in Eq.~(\ref{eq: early N yield}), $x_{{\rm id},N}^{3}$ is given below that equation, and $c$ is an order one constant. 
The time  $x_{\rm eq}$ at which the $N$ density is equal to the radiation energy density, $\rho_{R}=\pi^{2} g_* m_N^4/(30 x^{4})$, is given by
\begin{equation}
x_{\rm eq}=\frac{\pi^{4} g_* x_{{\rm id},N}^{3} }{10\, c^3 \, g_N }.
\end{equation}
For this to happen before decay, $x_{\rm eq}<x_{{\rm d},N}=\sqrt{H(m_{N})/\Gamma_{N}}$, we need
\begin{equation}
\frac{\Gamma_{N}}{H(m_{N})}>\left(\frac{\pi^{4}g_{*}}{5g_{N}c^{3}}\right)^{2}\approx45,
\end{equation}
 for $g_{*} \approx 10.75.$ and $c \approx 2.5$. But this is enough to cause $N$ to equilibrate
with the SM, i.e., there is no consistent choice of parameters for which $N$ is long-lived and dominates the energy density before its decay. 

Furthermore, if the $N$ lifetime is longer than ${\cal O}(1 \, {\rm s})$ it could impact Big Bang Nucleosynthesis (BBN). However, for weak scale heavy neutrinos this only occurs from extremely small values of $y$, 
\begin{equation}
\Gamma_{N}\simeq 10\, {\rm s}^{-1} \left(\frac{y}{10^{-13}}\right)^{2}\left(\frac{m_{N}}{300\rm{~GeV}}\right).
\end{equation}
For such small couplings DM is significantly underproduced and are thus outside our range of interest.

\subsection{Results}
\begin{figure}
\centering
\includegraphics[width=0.65\textwidth]{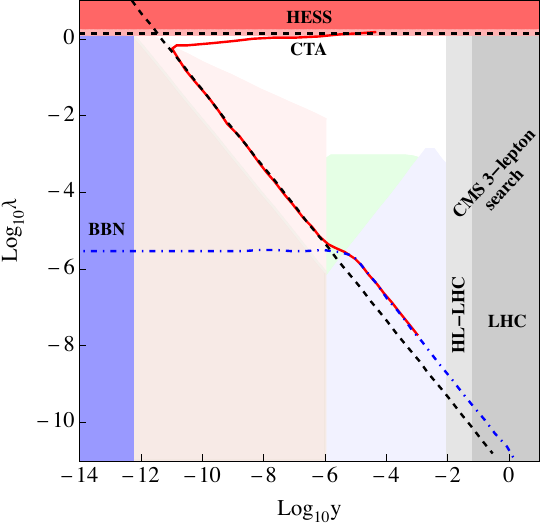}
\caption{{\it Constraints for $m_N=200\,{\rm{GeV}}, m_{\chi}=2\,{\rm{TeV}},m_{\phi}=3\,{\rm TeV}$}: 
The blue, dark red, and dark grey regions are excluded by Big Bang Nucleosynthesis (BBN), indirect detection searches (HESS), and collider searches (LHC). The light red and light grey shaded regions show the projected sensitivities of CTA and HL-LHC, respectively. The observed abundance of DM today is produced in our model for points that lie on the solid red line. This shows the result of numerically solving the Boltzmann equation.
The dashed lines are the approximate solutions for the observed relic density obtained in the text.
The qualitative regimes of the parameter space, as shown in Fig.~\ref{fig:paramsummary}, are indicated with faint colors.}
\label{fig:constraints}
\end{figure}

Fig.~\ref{fig:constraints} shows regions of the $(y,\lambda)$ parameter space excluded by various constraints for the choice of mass parameters $m_N=200\,{\rm{GeV}}, m_{\chi}=2\,{\rm{TeV}},m_{\phi}=3\,{\rm TeV}$. The excluded regions by BBN, indirect detection searches (HESS), and collider searches (LHC) are shaded in blue, dark red, and dark grey, respectively. The projected sensitivity of CTA is shown in light red and that of HL-LHC in light grey. The observed cosmological
abundance of DM is produced in our model for points that lie on the solid red line which is obtained by solving the Boltzmann equation numerically. The dashed lines are the approximate solutions obtained in the text.
\section{Discussion}
\label{sec:discussion}
 We have investigated the cosmological histories of a simple dark sector consisting of a complex scalar and a Dirac fermion (the DM candidate) which is populated from the SM bath through the neutrino portal. The portal Yukawa coupling between the sterile neutrino mediator and the SM is responsible for injecting energy into the dark sector, while the dark sector Yukawa coupling dictates how energy/momentum is redistributed among the dark sector species. For dark sector masses around the TeV scale and the mass ordering $m_N < m_\chi < m_\phi$, the thermal evolution of the dark sector and consequently the DM abundance falls into one of four qualitatively distinct phases, namely freeze-out, freeze-in, double freeze-in, and dark sector equilibration. Freeze-out corresponds to large values of coupling constants such that the dark sector establishes chemical equilibrium with the SM bath and subsequently DM freezes out from it. In the freeze-in regime, the sterile neutrino mediator is in equilibrium with the SM plasma due to a relatively large portal coupling, while the other dark sector particles never achieve equilibrium due to their minuscule dark coupling. Small values of both couplings lead to the double freeze-in phase wherein the sterile neutrino mediator never comes into thermal equilibrium with the SM plasma and DM is produced via the two-stage process of mediator freeze-in from the SM and the subsequent annihilations of the mediators to dark particles. Finally, if the mediator has a very weak coupling to the SM bath and a large dark interaction, the gradual leaking of energy from the SM bath to the dark sector is converted into a separate dark thermal bath with a temperature much smaller than the visible sector one due to dark sector equilibration.

By developing semi-analytic solutions of the integrated Boltzmann equations, we have provided estimates for the evolution of the energy density of the dark sector and the number densities of its constituents. These analytical estimates, validated by numerical solutions, have been used to delineate the parameter regions realizing the various cosmological phases. We have also highlighted how this model may be confronted with experimental and observational probes from cosmology, indirect detection, colliders,  precision electroweak data, and direct detection. 

As our study highlights, even a modest increase in complexity within the dark sector, whether through additional particle species or interactions, can lead to a much richer cosmological evolution, beyond the familiar paradigms of thermal freeze-out and freeze-in. Such scenarios not only allow for new mechanisms to determine the DM abundance but may also necessitate a rethinking of associated dark sector search strategies. These possibilities underscore the need for continued exploration of alternative dark sector cosmologies.

\acknowledgments
The work of B.B. is supported by the U.S. Department of Energy under grant No. DE–SC0007914.
The work of J.B. is supported by the National Science Foundation under Grant No.\ 2413017. 
The work of B.S.E. is supported in part by U.S. Department of Energy under grant No. DE-SC0022021. The work of D.M. is supported by Discovery Grants from the Natural Sciences and Engineering Research Council of Canada (NSERC). TRIUMF receives federal funding via a contribution agreement with the National Research Council (NRC) of Canada.

%
\appendix

\section{Thermally-Averaged Cross Sections for Particles at Different Temperatures}
\label{app:therm-avg}
In this Appendix, 
we discuss the thermal averages of cross sections and decay rates which appear throughout our analysis. 
We mainly concentrate on the reduction of the 
six-dimensional integrals appearing in thermal averages into a single integral, which then can be evaluated numerically. 
The results are well known in the literature for the case when the two colliding particles have the same temperature~\cite{Gondolo:1990dk}. For particles at different temperatures, one usually resorts to a non-relativistic expansion as in~\cite{Bringmann:2006mu}. 
Here, we present the full results, which to the best of our knowledge are new to the literature. To set the stage, it will be instructive to review the equal-temperature case first.

\subsection{Same-temperature case}
Consider the 2-to-2 reaction $a+b\rightarrow c+d$. If particles $a,b$ follow the equilibrium Maxwell-J$\ddot{\rm u}$ttner distribution~\cite{JttnerDasMG,Jttner1928DieRQ} 
with the same temperature $T$, then the thermal average of $\sigma v$ is given by: 
\begin{equation}
\langle(\sigma v)_{ab\rightarrow cd}\rangle
= \frac{1}{n_{a}^{\rm eq}n_{b}^{\rm eq}}\int \! \frac{ g_a d^{3}p_{a}}{(2\pi)^{3}}\frac{  g_b d^{3}p_{b}}{(2\pi)^{3}}\, e^{-E_{a}/T}e^{-E_{b}/T}\,\sigma v \,.
\end{equation}
Here $v$ is the Moeller velocity, and $n^{\rm eq}_i$ is the equilibrium number density for species $i$, with mass $m_i$ and $g_i$ internal degrees of freedom,
\begin{equation}
\label{eq:equil-number-density}
    n_i^{\rm eq}(T) = \int \frac{ g_i \, d^3 p_i}{(2\pi)^3} e^{-E_i/T} = \frac{  g_i  }{2 \pi^2} m_i^2 T K_2(m_i/T).
\end{equation}
We employ the convention that cross sections and decay widths are defined such that an average over all initial and sum over all final state degrees of freedom are performed. 
Before attempting to carry out the integrals, 
it is useful to examine some of their properties.
First, note that the quantity $\sigma v \, E_{a}E_{b}$ is Lorentz invariant and  
can be expressed in terms of the invariant squared centre-of-mass energy $s=(p_{a}+p_{b})^{2}$
as
\begin{equation}
\label{eq:sv-same-T-0}
\sigma v \, E_{a}E_{b} =\frac{1}{2}\sigma(s)\lambda^{1/2}(s,m^2_{a},m^2_{b})\,,
\end{equation}
where 
$\lambda(\hat x, \hat y, \hat z) \equiv \hat x^2 + \hat y^2 + \hat z^2 -2 \hat x \hat y - 2 \hat x \hat z - 2 \hat y \hat z$ is the familiar K\"{a}ll\'{e}n function.
This observation 
implies that the integral (hence the quantity
$\langle(\sigma v)_{ab\rightarrow cd}\rangle n_{a}^{\rm eq}n_{b}^{\rm eq}$) is  
Lorentz invariant if the distribution function is as well.  
In general, the distribution function is a Lorentz scalar~\cite{Liboff,Debbasch:2001voz}, but here we are evaluating
it in the cosmic frame. 
We conclude that we cannot use Lorentz invariance when evaluating the integrals. However, it is easy to see that the
integral is rotationally invariant, and hence spherical coordinates
become the natural coordinate choice. One can take advantage of the rotational invariance
to rotate $\boldsymbol{p_{b}}$ so that 
its $z$-axis coincides with the direction of $\boldsymbol{p_{a}}$, rendering 
the angle between the two vectors $\boldsymbol{p_{a}}$ and $\boldsymbol{p_{b}}$ the same as the polar angle $\theta$ of  $\boldsymbol{p_{b}}$. Since the integrand depends only on the
angle between the two vectors through its dependence on $s$,
one can integrate over all the other angles yielding a factor of $8\pi^{2}$. 
The measure then becomes (denoting the magnitude of the momentum vector by $p=\sqrt{\boldsymbol{p^{2}}}$)
\begin{equation}
\label{sv-measure}
d^{3}p_{a}\, d^{3}p_{b}=8\pi^{2}\, p_{a}^{2} \, dp_{a} \, p_{b}^{2} \, dp_{b} \,d\cos\theta 
= 8\pi^{2} \, p_{a}  E_{a} \, dE_{a} \, p_{b}  E_{b} \, dE_{b} \, d\cos\theta\,, 
\end{equation}
leading to
\begin{equation}
\label{eq:sv-same-T-1}
\langle(\sigma v)_{ab\rightarrow cd}\rangle \, n_{a}^{\rm eq}\, n_{b}^{\rm eq}
= \frac{ g_a g_b}{16\pi^{4}}\int dE_{a} \, dE_{b} \, d\cos\theta \, p_{a} \, p_{b} 
\, \sigma(s) \, \lambda^{1/2}(s,m^2_{a},m^2_{b}) \, e^{-(E_{a}+E_{b})/T}\,.
\end{equation}

The form of the integrand suggests a change of variables from $(\cos\theta,E_{a},E_{b})$
to 
$s=m_{a}^{2}+m_{b}^{2}-2E_{a}E_{b}-2p_{a}p_{b}\cos\theta$
and the combination $E_{+}=E_{a}+E_{b}$ that appears in the exponential.
One then chooses the combination $E_{-}=E_{a}-E_{b}$ that does not appear
in the integrand as the third variable, and thus obtains 
$dE_{+} \, dE_{-} \, ds = 4\,p_{a} \, p_{b} \, dE_{a} \, dE_{b} \, d\cos\theta$. Eq.~(\ref{eq:sv-same-T-1}) then becomes 
\begin{equation}
\label{eq:sv-same-T-2}
\langle(\sigma v)_{ab\rightarrow cd}\rangle \, n_{a}^{\rm eq}\, n_{b}^{\rm eq}
 = \frac{ g_a g_b}{64\pi^{4}}\int \! dE_{+}\, dE_{-}\, ds \, \sigma(s) \, \lambda^{1/2}(s,m^2_{a},m^2_{b})e^{-E_{+}/T}\,.
\end{equation}
The region of integration $E_{a}\geq m_{a}$, $E_{b}\geq m_{b}$, $\left|\cos\theta\right|\leq1$
corresponds to a region in the $(E_{+},E_{-},s)$ space. 
The $s$ limits are easiest to obtain: $s=(E_{a}+E_{b})^{2}-(\boldsymbol{p_{a}+p_{b}})^{2}$
implies that $s^{\rm  min}= M^2 \equiv (m_{a}+m_{b})^{2}$ 
(by using $E^{\rm min}=m$, $p^{\rm min}=0$),
while $s^{\rm max}\rightarrow\infty$ (since $E\geq p$). 
This definition of $s$ also helps in determining the $E_{+}$ limits: 
$E_{+}^{2}=s+(\boldsymbol{p_{a}}+\boldsymbol{p_{b}})^{2}$
leads to $E_{+}^{\rm min}=\sqrt{s}$ ($p=0$) and $E_{+}^{\rm max}\rightarrow\infty$.
Finally, the $E_{-}$ limits involve a fair bit of algebra: one starts
from the condition $\left|\cos\theta\right|\leq1$ expressed as $\left|\frac{s-m_{a}^{2}-m_{b}^{2}-2E_{a}E_{b}}{-2p_{a}p_{b}}\right|\leq1$,
writes $E_{a,b}$ and $p_{a,b}$ in terms of $E_{\pm}$, and solves for the $E_{-}$ boundaries. Putting everything together, 
the integration region is defined by the following:
\begin{align}
\label{eq:limits-same-T}
&  s \geq M^{2} = (m_a+ m_b)^2\,, \quad\quad\quad  
E_{+} \geq\sqrt{s}\,,  \quad\quad\quad   
E_{-}^{\rm min} \leq E_- \leq E_{-}^{\rm max}\,,\\
& \quad\quad\quad    E_{-}^{\rm max(min)} = \frac{m_{a}^{2}-m_{b}^{2}}{s}E_{+}\pm\frac{\lambda^{1/2}(s,m^2_{a},m^2_{b})}{s}\sqrt{E_{+}^{2}-s}
\,. \nonumber
\end{align}
One can now carry out the integrals in succession: the integrand is
independent of $E_{-}$, and the $E_{-}$ integral thus gives 
$E_{-}^{\rm max}-E_{-}^{\rm min}=(2/s)\lambda^{1/2}(s,m^2_{a},m^2_{b})\sqrt{E_{+}^{2}-s}$.
Eq.~(\ref{eq:sv-same-T-2}) becomes
\begin{equation}
\label{eq:sv-same-T-3}
\langle(\sigma v)_{ab\rightarrow cd}\rangle \, n_{a}^{\rm eq} \, n_{b}^{\rm eq} \, 
= \frac{ g_a g_b}{32\pi^{4}}\int_{M^{2}}^{\infty}\frac{ds}{s}\,\sigma(s)\,\lambda(s,m^2_{a},m^2_{b})
\int_{\sqrt{s}}^{\infty} dE_{+} \, \sqrt{E_{+}^{2}-s} \,\, e^{-E_{+}/T}\,,
\end{equation}
or upon carrying out the $E_{+}$ integral,
\begin{equation}
\label{eq:sv-same-T-4}
\langle(\sigma v)_{ab\rightarrow cd}\rangle \, n_{a}^{\rm eq} \, n_{b}^{\rm eq} \, 
= \frac{ g_a g_b \,T}{32\pi^{4}} 
\int_{M^{2}}^{\infty}\frac{ds}{\sqrt{s}} \,
\sigma(s)
\,\lambda(s,m^2_{a},m^2_{b}) \,
K_{1}\left[\frac{\sqrt{s}}{T}\right]\,,
\end{equation}
where we have used the modified Bessel function integral representation (see Ref.~\cite{gradshteyn2007})
\begin{equation} \label{eq: Bessel int rep}
K_{\nu}(z)=\frac{\sqrt{\pi}}{\Gamma(\nu+\frac{1}{2})}\left(\frac{z}{2}\right)^{\nu} \int_{0}^{\infty} dt \, e^{-z\cosh t} \, \sinh^{2\nu}t \,.
\end{equation}
Hence, we have reduced the six-dimensional momentum integral into a
single integral over the squared CM energy given by Eq.~(\ref{eq:sv-same-T-4}). The thermal average of $\sigma v E_a$ 
can be similarly reduced:
\begin{equation}
\label{eq:svE-same-T-2}
\langle(\sigma v E_a)_{ab\rightarrow cd}\rangle \, n_{a}^{\rm eq} \, n_{b}^{\rm eq} 
=\frac{ g_a g_b \, T}{32\pi^{4}}\int_{M^{2}}^{\infty}\frac{ds}{\sqrt{s}} \, \sigma(s)
\,\lambda(s,m^2_{a},m^2_{b})\,
\left[\frac{s+m_{a}^{2}-m_{b}^{2}}{2\sqrt{s}}\right]K_{2}\left[\frac{\sqrt{s}}{T}\right].
\end{equation}
Note that Eq.~(\ref{eq:svE-same-T-2}) is almost the same as Eq.~(\ref{eq:sv-same-T-4}) except for the additional factor
of energy of $a$ expressed in terms of the total CM energy, $E_{a}=\frac{s+m_{a}^{2}-m_{b}^{2}}{2\sqrt{s}}$, 
and 
the exchange of $K_{1}$ for $K_{2}$. 
As a result, one can readily obtain the integral for $\langle(\sigma v(E_{a}+E_{b}))_{ab\rightarrow cd}\rangle$:
\begin{equation}
\label{eq:svE-same-T-3}
\langle(\sigma v(E_{a}+E_{b}))_{ab\rightarrow cd}\rangle  \, n_{a}^{\rm eq} \, n_{b}^{\rm eq} 
=\frac{ g_a g_b \, T}{32\pi^{4}}\int_{M^{2}}^{\infty} \, ds \, \sigma(s) \,
\,\lambda(s,m^2_{a},m^2_{b}) \,
K_{2}\left[\frac{\sqrt{s}}{T}\right]\,.
\end{equation}

Finally, for a decay process $a\rightarrow b+c$, where the particle $a$ has temperature $T$, the thermally-averaged decay rate integrals are straightforward to evaluate and we only quote the well-known results:
\begin{align}
\label{eq:decay-same-T}
    \langle \Gamma_{a\rightarrow b c} \rangle & = \Gamma_{a\rightarrow b c} \, \bigg\langle \frac{m_a}{E_a} \bigg\rangle  
    = \Gamma_{a\rightarrow b c}\, \frac{K_1(m_a/T)}{K_2(m_a/T)}\, , \\
      \langle \Gamma_{a\rightarrow b c} E_a \rangle & = m_a \Gamma_{a\rightarrow b c}\, , \nonumber
\end{align}
where $\Gamma_{a\rightarrow b c}$ is the partial decay width in vacuum.

\subsection{Different temperature case: }

Here we assume that the two particles $a$ and $b$ have Maxwell-J$\ddot{\rm u}$ttner distributions with different temperatures $T_{a}$ and $T_{b}$, respectively. The thermal average of 
$\sigma v$ is
\begin{equation}
\label{eq:sv-different-T-1}
\langle(\sigma v)_{ab\rightarrow cd}\rangle= 
\frac{1}{ n_{a}^{\rm eq} \, n_{b}^{\rm eq} } \int\frac{ g_a   d^{3}p_{a}}{(2\pi)^{3}}\frac{ g_b d^{3}p_{b}}{(2\pi)^{3}}
\, e^{-E_{a}/T_{a}} e^{-E_{b}/T_{b}} \, \sigma v\,.
\end{equation}
The different temperatures do not affect the rotational invariance
of the integrand and one can readily use Eq.~(\ref{sv-measure}) to get
\begin{equation}
\label{eq:sv-different-T-2}
\langle(\sigma v)_{ab\rightarrow cd}\rangle  \, n_{a}^{\rm eq} \, n_{b}^{\rm eq}
= \frac{ g_a g_b }{16\pi^{4}}\int dE_{a} \, dE_{b} \, d\cos\theta \, p_{a} \, p_{b} \,  \sigma(s) \, \lambda^{1/2}(s,m^2_{a},m^2_{b}) \, e^{-\left(E_{a}/T_{a}+E_{b}/T_{b} \right)}\,.
\end{equation}
The $s-\cos\theta$ relation still suggests the change of variables
$\cos\theta\rightarrow s$, but the simple combination $E_{+}=E_{a}+E_{b}$
does not appear in the integrand in this case. Looking back to the same temperature
case above  for inspiration, we see that the key to simplifying the integral
was obtaining an integrand that is independent of $E_{-}$ by defining
the combination in the exponential to be the variable $E_{+}$. Following
this path, we define our new $E_{+}$ variable to be
\begin{equation}
\label{eq:EP-different-T}
\frac{E_{+}}{T}=\frac{E_{a}}{T_{a}}+\frac{E_{b}}{T_{b}}\,,
\end{equation}
 while $E_{-}$ is the other combination,
\begin{equation}
\label{eq:EM-different-T}
\frac{E_{-}}{T}=\frac{E_{a}}{T_{a}}-\frac{E_{b}}{T_{b}}\,,
\end{equation}
 where $T$ is an arbitrary reference ``temperature'' 
 (up to the fact that in the limit $T_{a}=T_{b}$, it is perfectly determined to be $T=T_{a}=T_{b}$)
inserted only for dimensional purposes, and therefore one expects the final result to be independent of it. These two equations, (\ref{eq:EP-different-T},\ref{eq:EM-different-T}), together
with the relation $s=m_{a}^{2}+m_{b}^{2}+2E_{a}E_{b}-2p_{a}p_{b}\cos\theta$,
constitute our variable transformation. The Jacobian of the transformation
is easily obtained:
\begin{equation}
ds \, dE_{+} \, dE_{-} = 4 \, \frac{T^{2}}{T_{a} \, T_{b}} \, p_{a} \, p_{b} \, dE_{a} \, dE_{b} \, d\cos\theta\,.
\end{equation}
Using this in Eq.~(\ref{eq:sv-different-T-2}), the equivalent of Eq.~(\ref{eq:sv-same-T-2}) is
\begin{equation}
\label{eq:sv-different-T-3}
\langle(\sigma v)_{ab\rightarrow cd}\rangle  \, n_{a}^{\rm eq} \, n_{b}^{\rm eq}  = \frac{ g_a g_b  }{64\pi^{4}}\frac{T_{a}T_{b}}{T^{2}}\int_{M^{2}}^{\infty} \, ds \, \sigma(s) \, \lambda^{1/2}(s,m^2_{a},m^2_{b}) \, \int_{E_{+}^{\rm min}}^{\infty} \, dE_{+} \, e^{-E_{+}/T} \, \int_{E_{-}^{\rm min}}^{E_{-}^{\rm max}} \, dE_{-}\, .
\end{equation}
The
integration boundaries can be worked out similarly to the same temperature case considered above, although it involves a bit of tedious algebra for the $E_{-}$ limits. We therefore quote only the results:
\begin{align}
\label{eq:limits-different-T}
&  \quad\quad   s \geq M^{2} = (m_a+m_b)^2\,, \quad\quad\quad  
E_{+}^{\rm min} \leq E_{+}  \leq \infty \,,  \quad\quad\quad  
E_{-}^{\rm min} \leq E_- \leq E_{-}^{\rm max}\,,   \\
& \quad\quad\quad\quad\quad\quad\quad\quad\quad E_{+}^{\rm min}=T\left(\frac{s-m_{a}^{2}-m_{b}^{2}}{T_{a}T_{b}}+\frac{m_{a}^{2}}{T_{a}^{2}}+\frac{m_{b}^{2}}{T_{b}^{2}}\right)^{1/2}\,, \nonumber \\
& \quad\quad\quad E_{-}^{\rm max(min)} = \frac{T^2}{(E_+^{\rm min})^2} \left[ \left(\frac{m_{a}^{2}}{T_{a}^{2}}-\frac{m_{b}^{2}}{T_{b}^{2}}\right) E_+
\pm \frac{\lambda^{1/2}(s,m^2_{a},m^2_{b})}{T_{a}T_{b}} \sqrt{E_+^2+(E_+^{\rm min})^2} \right]\,. \nonumber
\end{align}
It is now straightforward to carry out the integrals. For the $E_-$ integral, we have
\begin{align}
\int_{E_{-}^{\rm min}}^{E_{-}^{\rm max}}dE_{-} = 2 \, \frac{T^{2}}{(E_{+}^{\rm min})^{2}} \, 
\frac{\lambda^{1/2}(s,m^2_{a},m^2_{b})}{T_{a}T_{b}} \,
\sqrt{E_{+}^{2}-(E_{+}^{\rm min})^{2}}\,.
\end{align}
The $E_{+}$ integral is the same as that for the equal temperature case with the replacement $\sqrt{s}\rightarrow E_{+}^{\rm min}$:
\begin{align}
\int_{E_{+}^{\rm min}}^{\infty}dE_{+} \, \sqrt{E_{+}^{2}-(E_{+}^{\rm min})^{2}} \,\, e^{-E_{+}/T}  = 
E_{+}^{\rm min} \, T \, K_{1}\left[\frac{E_{+}^{\rm min}}{T}\right].
\end{align}
Putting everything together, Eq.~(\ref{eq:sv-different-T-3}) becomes:
\begin{equation}
\label{eq:sv-different-T-4}
\langle(\sigma v)_{ab\rightarrow cd}\rangle  \, n_{a}^{\rm eq} \, n_{b}^{\rm eq}   = \frac{ g_a g_b \, T}{32\pi^{4}}
\int_{M^{2}}^{\infty}\frac{ds}{E_{+}^{\rm min}}\,\sigma(s) \,
\lambda(s,m^2_{a},m^2_{b}) \,
K_{1}\left[\frac{E_{+}^{\rm min}}{T}\right]\,,
\end{equation}
which is identical to Eq.~(\ref{eq:sv-same-T-4}) -- its equivalent for the equal
temperature case -- except for the replacement $\sqrt{s}\rightarrow E_{+}^{\rm min}$ in the appropriate positions. We see that the final result (\ref{eq:sv-different-T-4}) does not depend on the arbitrary reference ``temperature'' $T$ as promised, owing to the fact that 
$E_{+}^{\rm min}/T$ is independent of $T$ (see Eq.~(\ref{eq:limits-different-T})). Note that in the limit $T_a = T_b = T$ we reproduce Eq.~(\ref{eq:sv-same-T-4}).

In a similar way, one can carry out the energy density integral to get 
\begin{equation}
\label{eq:sv-different-T-5}
\begin{split}\langle(\sigma vE_a)_{ab\rightarrow cd}\rangle \, n_{a}^{\rm eq} \, n_{b}^{\rm eq} = \frac{ g_a g_b  T_{a}}{64\pi^{4}}\int_{M^{2}}^{\infty}
 ds \,  \sigma(s) \, \lambda(s,m_a^2,m_b^2) \left[1+\frac{T^2}{(E_{+}^{\rm min})^2} \left( \frac{m_a^2}{T_a^2} - \frac{m_b^2}{T_b^2} \right)   \right] 
K_{2}\left[\frac{E_{+}^{\rm min}}{T}\right],
\end{split}
\end{equation}
where $T_{a}$ appears out front because the energy factor appearing in the average is
$E_{a}$. 

For decay processes, the results for the case with particle $a$ at a different temperature $T'$ than the SM bath temperature $T$ can be obtained from Eq.~(\ref{eq:decay-same-T}) by replacing $T\rightarrow T'$.

\subsection{Cross Section and Decay Width Formulae and Approximations}
Here we collect the formulae for the cross sections and decay widths for the various processes appearing in the thermal averages that enter in the Boltzmann equations. 
We also present approximate formulae for thermal averages in the 
relativistic and non-relativistic regimes (for only the $T_{a}=T_{b}$ case), 
which are used in our semi-analytic analyses in various regions of parameter space. Before considering specific processes, let us note that $N_c = 3$ ($N_L = 2$)  are color factors associated with $SU(3)_c$ ($SU(2)_L$) groups. 
Also, the factors $g_N = 2$, $g_\chi = 2$, $g_\phi = 1$, $g_H = 2$, $g_L = 4$, $g_t = 6$, and $g_Q = 12$ count the internal degrees of freedom (polarizations + colors) of the various species. 

\begin{itemize}
\item $N$ interactions with SM: (inverse) decays and scatterings with top quarks.
\end{itemize}
The dominant interactions of $N$ with the SM occur through the (inverse) decays ($N \leftrightarrow \overline{H} + L$) 
and through scatterings of SM top quarks ($t + \overline{Q} \leftrightarrow N + \overline{L}$ plus crossed reactions). 
The thermal average of this rate including both decays and scatterings is 
\begin{equation}
\label{eq:N-rate}
\langle\Gamma_{N,{\rm tot}}\rangle_{T'}=\Gamma_{N}\, \frac{K_{1}(m_{N}/T')}{K_{2}(m_{N}/T')}+\langle(\sigma v)_{N,{\rm SM}}\rangle_{T'} \,n_{\rm SM}.
\end{equation}
Here the decay width $\Gamma_N  \equiv \Gamma_{N\rightarrow \overline{H} + L}$ is given by 
\begin{equation}
\Gamma_N = \frac{N_L}{g_N} \frac{y^2 m_N}{16 \pi} \left(1-\frac{m_H^2}{m_N^2}\right)^2.
\end{equation}
The second term in (\ref{eq:N-rate}) is obtained from Eqs.~(\ref{eq:sv-different-T-4},\ref{eq:equil-number-density}) with $T_{a}=T$, $m_{a}=0$, $n_{a}^{\rm eq}=n_{\rm SM} = g_{\rm SM} T^3/\pi^2$ with $g_{\rm SM} = 1$ (equilibrium phase space distribution for a single massless SM degree of freedom),
$T_{b}=T'$, $m_{2}=m_{N}$, and may be written as 
\begin{equation}
\langle(\sigma v)_{N,{\rm SM}}\rangle_{T'} = \frac{1}{n_{\rm SM} n_N^{\rm eq}(T')}  \frac{\sqrt{T T'}}{32 \pi^4} \int_{m_N^2}^\infty \!\! ds \, \sigma_{N,{\rm SM}}
\frac{(s-m_N^2)^2}{\sqrt{s-m_N^2(1-T/T')}} K_1 \left[\sqrt{\frac{s-m_N^2(1-T/T')}{T T'}}  \right].
\end{equation}
Here we have defined the total effective cross section for $N$ scattering with SM degrees of freedom, $\sigma_{N, {\rm SM}} \equiv g_N g_L \, \sigma_{N\overline L \rightarrow t \overline Q}  + g_N g_Q \, \sigma_{N Q \rightarrow L t} + g_N g_t \, \sigma_{N\overline t \rightarrow L \overline Q}$, which is given by
\begin{align}
\sigma_{N, {\rm SM}} =  N_c N_L \frac{y^2 y_t^2}{16 \pi} \left\{ \frac{s}{(s- m_H^2)^2}\!  + \!  \frac{2}{s- m_N^2} 
 \left[ 1 \! -\! \frac{m_N^2 -  m_H^2}{ s- m_N^2 + m_H^2 }\!  + \! \frac{m_N^2 - 2 m_H^2 }{s- m_N^2} \ln\left(\!1 \!+ \! \frac{s \!- \!  m_N^2}{m_H^2} \!  \right)  \right]    \right\}.
\end{align}
In a similar fashion, we may define the rate $\langle\Gamma_{N,{\rm tot}} E_N \rangle_{T'}$ from Eq.~(\ref{eq:sv-different-T-5}). We also note the approximate expressions for the thermal averages at high temperatures, which are used in our analysis in Section~\ref{sec:cosmo}:
\begin{align}
\langle(\sigma v)_{N,{\rm SM}}\rangle_{T}^{\rm rel}& \simeq \frac{ 3N_c N_L \, y_{t}^{2} \, y^{2}}{128\pi g_{N}g_{\rm SM}T^{2}}, \\
\langle(\sigma vE_{N})_{N, {\rm SM} }\rangle_{T}^{\rm rel} & \simeq \frac{3 N_c  N_L \, y_{t}^{2} \, y^{2}}{64\pi g_{N}g_{\rm SM}T}.
\end{align}

\begin{itemize}
\item $\chi+\overline{\phi}\rightarrow L+\overline{H}$ 
\end{itemize}
The cross section $ \sigma_{\chi \phi LH} \equiv \sigma_{\chi+\overline{\phi}\rightarrow L+\overline{H}}$ is given by
\begin{equation}
    \sigma_{\chi \phi LH} = \frac{N_L}{g_\chi g_\phi} \frac{y^2 \lambda^2}{32 \pi s} 
    \frac{ \lambda^{3/2}(1,m_H^2/s,0)}{\lambda^{1/2}(1,m_\chi^2/s,m_\phi^2/s)}
    \left[\frac{(s+m_{\chi}^{2}-m_{\phi}^{2})(s+m_{N}^{2})+4m_{\chi}m_{N}s}{(s-m_{N}^{2})^{2}}\right].
\end{equation}
We also provide approximations for the thermal averages in different regimes, which are used in our analysis in Section~\ref{sec:cosmo}:
\begin{align}
\langle(\sigma v)_{\chi\phi LH}\rangle_{T}^{\rm rel} & \simeq\frac{N_L y^{2}\lambda^{2}}{256\pi g_{\chi}g_{\phi}T^{2}} , \\
\langle(\sigma v)_{\chi\phi LH}\rangle^{\rm Nrel} & \simeq\frac{N_L y^{2}\lambda^{2}(m_{\phi}+m_{\chi})}{32\pi g_{\chi}g_{\phi}m_{\phi}(m_{\phi}+m_{\chi}-m_{N})^{2}} ,\\
\langle(\sigma v(E_{\chi}+E_{\phi}))_{\chi\phi LH}\rangle_{T}^{\rm rel} & \simeq\frac{N_L y^{2}\lambda^{2}}{64\pi g_{\chi}g_{\phi}T} ,\\
\langle(\sigma v(E_{\chi}+E_{\phi}))_{\chi\phi LH}\rangle^{\rm Nrel} & \simeq\frac{N_L y^{2}\lambda^{2}(m_{\phi}+m_{\chi})^{2}}{32\pi g_{\chi}g_{\phi}m_{\phi}(m_{\phi}+m_{\chi}-m_{N})^{2}} .
\end{align}

\begin{itemize}
\item $\chi+\overline{\chi}\rightarrow N+\overline{N}$ 
\end{itemize}
The cross section $ \sigma_{\chi \chi N N} \equiv \sigma_{\chi+\overline{\chi}\rightarrow N+\overline{N}}$ is given by
\begin{align}
\label{eq:xsec-NNXX}
\sigma_{\chi \chi  N  N} & =  \frac{1}{g_\chi^2} \frac{\lambda^4}{4 \pi} \frac{\beta_N}{ \beta_\chi} \frac{1}{s} \bigg\{1 
+ \frac{ [m_\phi^2  - (m_\chi + m_N)^2 ]^2}{m_\phi^2  s + \lambda(m_\phi^2, m_\chi^2, m_N^2) } \\ 
& ~~~~ -  \frac{ 2 [m_\phi^2  - (m_\chi + m_N)^2 ]}{s \beta_\chi \beta_N } \ln\left[ \frac{   s(1+\beta_\chi \beta_N ) + 2 (m_\phi^2 - m_\chi^2 - m_N^2) }{  s(1 - \beta_\chi \beta_N ) + 2 (m_\phi^2 - m_\chi^2 - m_N^2)  }   \right]
 \bigg\}.\nonumber
 \end{align}
 Here, $\beta_i = \lambda^{1/2}(1,m_i^2/s,m_i^2/s) = (1-4 m_i^2/s)^{1/2}$.
The thermal averages in relativistic and non-relativistic regimes, for use in Section~\ref{sec:cosmo}, are, respectively
\begin{align}
\langle(\sigma v)_{\chi\chi NN}\rangle_{T^{'}}^{\rm rel} & \simeq\frac{\lambda^{4}}{32\pi g_{\chi}^{2}T{}^{'2}}, \\
\langle(\sigma v)_{\chi\chi  NN}\rangle^{\rm Nrel} & \simeq\frac{\lambda^{4}}{2\pi g_{\chi}^{2}}\left(\frac{m_{\chi}+m_{N}}{m_{\phi}^{2}+m_{\chi}^{2}-m_{N}^{2}}\right)^{2} \left(1- \frac{m_N^2}{m_\chi^2}  \right)^{1/2}.
\end{align}

\begin{itemize}
\item $\phi+\overline \phi \rightarrow N+\overline{N}$ 
\end{itemize}
The cross section $\sigma_{\phi \phi N N} \equiv \sigma_{\phi+\overline \phi \rightarrow N+\overline{N}}$ is given by
\begin{align} 
\label{eq:xsec-NNphiphi} 
\sigma_{\phi \phi N N} & =  \frac{1}{g_\phi^2} \frac{\lambda^4}{8 \pi} \frac{\beta_N}{ \beta_\phi} \frac{1}{s} \bigg\{ \! - 1 
-  \frac{ [m_\phi^2  - (m_\chi + m_N)^2 ]^2}{m_\chi^2  s + \lambda(m_\phi^2, m_\chi^2, m_N^2) }  \\ 
& ~~~~ +  \frac{\{ s -  2 [m_\phi^2  - (m_\chi + m_N)^2 ] \}}{s \beta_\phi \beta_N } \ln\left[ \frac{   s(1+\beta_\phi \beta_N ) - 2 (m_\phi^2 - m_\chi^2 + m_N^2) }{  s(1 - \beta_\phi \beta_N ) - 2 (m_\phi^2 - m_\chi^2 + m_N^2)  }   \right]
 \bigg\}. \nonumber
\end{align}
The thermal averages in relativistic and non-relativistic regimes, for use in Section~\ref{sec:cosmo}, are, respectively
\begin{align}
\langle(\sigma v)_{\phi\phi NN}\rangle_{T'}^{\rm rel} & \simeq\frac{\lambda^{4}}{32 \pi g_{\phi}^{2}T{}^{'2}}\left[\ln\left(\frac{2 T^{'}}{m_{\chi}}\right)-\gamma\right],\\
\langle(\sigma v)_{\phi\phi NN}\rangle^{\rm Nrel}& \simeq \frac{\lambda^{4}}{4\pi g_{\phi}^{2}}\left(\frac{m_{\chi}+m_{N}}{m_{\phi}^{2}+m_{\chi}^{2}-m_{N}^{2}}\right)^{2} \left(1-\frac{m_N^2}{m_\phi^2}\right)^{3/2}.
\end{align}

\begin{itemize}
\item $\phi+\overline \phi \rightarrow \chi+\overline{\chi}$ 
\end{itemize}
The cross section $\sigma_{\phi \phi \chi \chi} \equiv \sigma_{\phi+\overline \phi \rightarrow \chi +\overline{\chi}}$ is given by
\begin{align}
\label{eq:xsec-XXphiphi}
\sigma_{\phi \overline \phi \rightarrow \chi \overline \chi} & =  \frac{1}{g_\phi^2} \frac{\lambda^4}{8 \pi} \frac{\beta_\chi}{ \beta_\phi} \frac{1}{s} \bigg\{ \! - 1 
-  \frac{ [m_\phi^2  - (m_\chi + m_N)^2 ]^2}{m_N^2  s + \lambda(m_\phi^2, m_\chi^2, m_N^2) } \\ 
& ~~~~ +  \frac{\{ s -  2 [m_\phi^2  - (m_\chi + m_N)^2 ] \}}{s \beta_\phi \beta_\chi } \ln\left[ \frac{   s(1+\beta_\phi \beta_\chi ) - 2 (m_\phi^2 + m_\chi^2 - m_N^2) }{  s(1 - \beta_\phi \beta_\chi ) - 2 (m_\phi^2 + m_\chi^2 - m_N^2)  }   \right]
 \bigg\}. \nonumber
 \end{align}

\begin{itemize}
\item  $\phi\leftrightarrow\chi +\overline{N}$
\end{itemize}
The partial decay width $\Gamma_{\phi} \equiv \Gamma_{\phi \rightarrow \chi +\overline{N}}$ is given by
\begin{equation}
\label{eq:Gamma-phi}
\Gamma_{\phi}   =   \frac{1}{g_\phi}  \, \frac{\lambda^2 m_\phi}{8\pi}  \left[    1- \left( \frac{   m_\chi + m_N }{ m_\phi } \right)^2    \right] \lambda^{1/2}\left(1, \frac{m_\chi^2}{m_\phi^2}, \frac{m_N^2}{m_\phi^2}   \right).
\end{equation}

\bibliographystyle{utphys}
\bibliography{Nportal}

\end{document}